\documentclass[prl,letterpaper,aps,10pt,twocolumn,floatfix,showpacs]{revtex4-2}

\usepackage{graphicx}
\usepackage{dcolumn}
\usepackage{bm}
\usepackage{color,soul}
\usepackage{amsmath}

\begin{document}

\title{Phase Symmetry Breaking of Counterpropagating Light in Microresonators\\ for Switches and Logic Gates}
\author{Alekhya \surname{Ghosh$^{\;1,2,\dagger}$}}
\author{Arghadeep \surname{Pal}$^{\;1,2,\dagger}$}
\author{Shuangyou \surname{Zhang}$^{\;1}$}
\author{Lewis \surname{Hill$^{\;1}$}}
\author{Toby \surname{Bi}$^{\;1,2}$}
\author{Pascal \surname{Del'Haye$^{\;1,2,*}$}}
\affiliation{$^1$Max Planck Institute for the Science of Light, Staudtstra{\ss}e 2,
D-91058 Erlangen, Germany\\$^2$Department of Physics, Friedrich Alexander University Erlangen-Nuremberg, D-91058 Erlangen, Germany}
\affiliation{$^\dagger$These authors contributed equally\\$^*$pascal.delhaye$@$mpl.mpg.de}

\begin{abstract}
The rapidly growing field of integrated photonics is enabling a large number of novel devices for optical data processing, neuromorphic computing and circuits for quantum photonics. While many photonic devices are based on linear optics, nonlinear responses at low threshold power are of high interest for optical switching and computing. In the case of counterpropagating light, nonlinear interactions can be utilized for chip-based isolators and logic gates. In our work we find a symmetry breaking of the phases of counterpropagating light waves in high-Q ring resonators. This abrupt change in the phases can be used for optical switches and logic gates. In addition to our experimental results, we provide theoretical models that describe the phase symmetry breaking of counterpropagating light in ring resonators.
\end{abstract}

\maketitle
Recent years have seen huge progress in the use of photonics for data processing, neural networks and quantum circuits. However, it is still challenging to add nonlinear responses to all-optical circuits. One solution is the use of optical-to-electronic data conversion, with the drawback of adding complexity and increases latency. In terms of all-optical systems, there has been progress to realize optical switches and logic gates using fibers~\cite{Gan:15, PhysRevA.79.030303}, photonic crystal waveguides~\cite{tanabe2005fast, kim2013quantum}, plasmonic waveguides~\cite{guo2022femtojoule, ono2020ultrafast, Hoessbacher:14,wei2011cascaded}, and microresonators~\cite{almeida2004all,raja2021ultrafast,Xu:11,van2002all,DelBino:18,godbole2016all,sethi2014all,qiu2020high,hill2004fast,maes2006switching,hamel2015spontaneous}. Compared to the rest, microresonators have gained much attention due to their well-established and simple fabrication methods along with low threshold power for driving optical nonlinearities.\\
\indent In this work, we utilize the phase response of optical fields undergoing spontaneous symmetry breaking (SSB) of intensities in microresonators to build an all-optical switch. We further propose designs of all-optical XOR and universal NAND gates realizable in integrated systems. SSB of counter-propagating~\cite{KAPLAN1982229,woodley2018universal, DelBino2017} or copropagating light fields with orthogonal polarizations~\cite{PhysRevLett.122.013905, garbin2020asymmetric, hill2024symmetry, campbell2024frequency} in microresonators can be used for applications ranging from isolators to logic-gates and gyroscopes~\cite{DelBino:18, DelBino:21, Silver:21, Moroney:20, white2023integrated, Moroney2022, del2021optical,yoshiki2014all}. Recent research has also predicted multi-level SSBs~\cite{hill2023multi,ghosh2023four,ghosh2024controlled,pal2024linear}. All this previous research investigated the SSB of intensities of the optical fields without taking into account the corresponding phase effects. Compared to previous work, our model does not need any optical power bias in any direction and is robust to large laser fluctuations and fabrication errors.\\
\begin{figure}[h!]
\includegraphics[width=1\columnwidth]{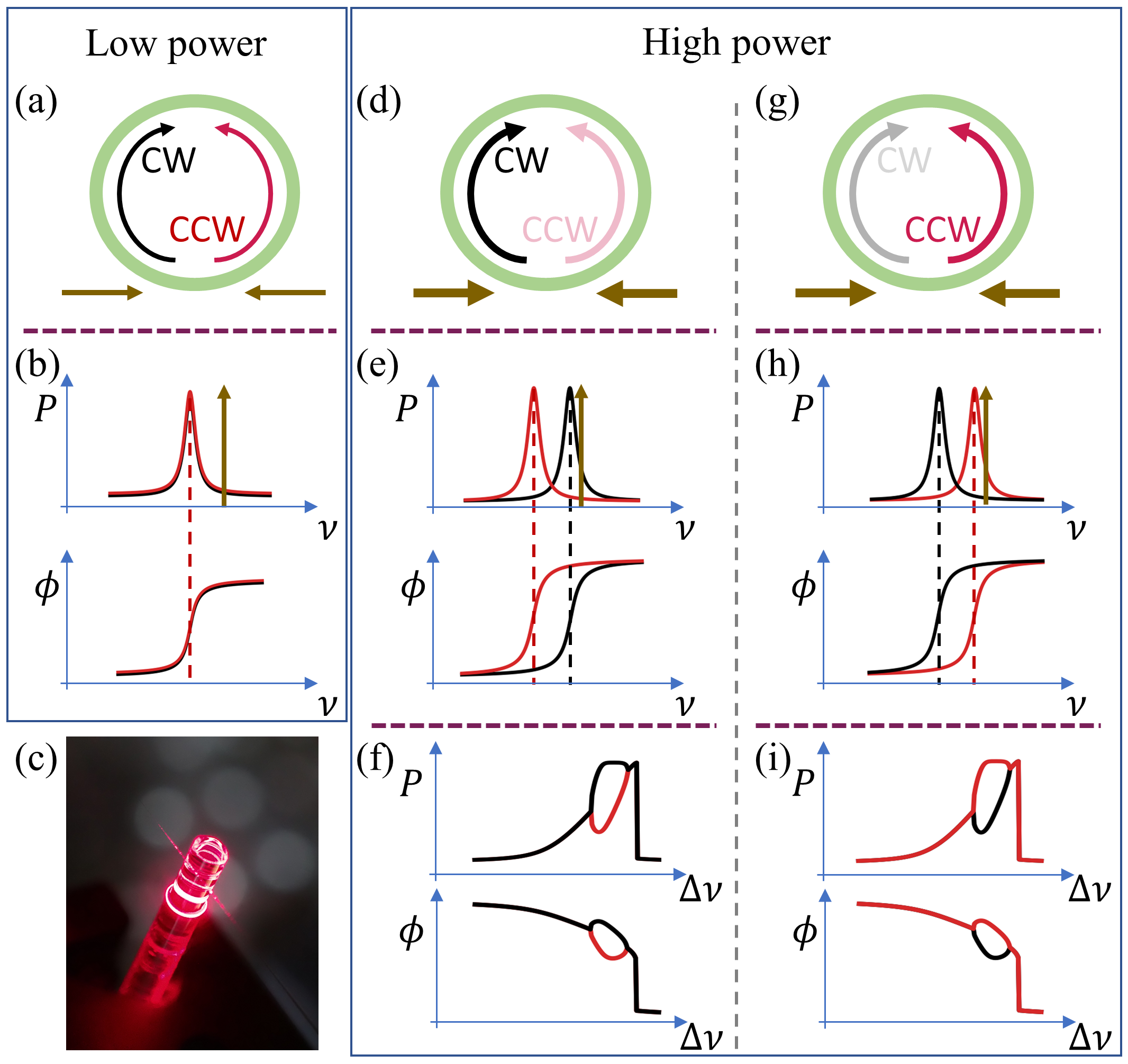}
\caption {\textit{Spontaneous symmetry breaking (SSB) of intensities and phases.} (a) At low input power, when the resonator is pumped bi-directionally, the intracavity power and phase profiles of the two circulating fields remain degenerate as shown in panel (b). (c) Photograph of the microrod resonator coupled to a tapered fiber that is used in the experiments. At high input power, the system drives into one of two states: (d) or (g), with more power circulating either in clockwise or counterclockwise direction. Panels (e) and (h) present the corresponding Kerr-induced resonance shifts. Panels (f) and (i) depict the SSB of both the propagating fields’ power and phase with varying laser detuning corresponding to (d) and (g) respectively. The clockwise (CW) and counter-clockwise (CCW) circulating directions are indicated by red and black arrows (and lines). In panels (a, d, g) the widths of the arrows correspond to intensities, solid and transparent arrows represent the dominant and suppressed fields after SSB, respectively.
}
\label{toyModel}
\end{figure}
\indent Figure~\ref{toyModel} illustrates the symmetry breaking between two counter-propagating fields in a bidirectionally pumped microresonator (Fig.~\ref{toyModel}(a)). At low input powers, when two counterpropagating input lasers approach the cavity resonances from the higher frequency sides, both directions see overlapping Lorentzian power profiles, as shown in the upper panel of Fig.~\ref{toyModel}(b). The corresponding phase profiles are depicted in the lower panel of Fig.~\ref{toyModel}(b). Panels (d-g) present the SSB mechanism at high power. After a certain threshold of input power~\cite{DelBino2017,woodley2018universal}, when the input lasers from the two directions approach the clockwise and counterclockwise resonances, an infinitesimal power imbalance between the two directions (caused by noise within the system), gets rapidly amplified, leading to SSB. This amplification of imbalance is fueled by unequal self- and cross-phase modulation strengths~\cite{hill2020effects}. The direction that receives slightly higher power due to random fluctuations pushes the counterpropagating resonance away from the input laser due to strong cross-phase modulation.\\
\indent In the upper panel of Fig.~\ref{toyModel}(e), the clockwise (CW) direction (black line) has pushed the counter-clockwise (CCW) direction (red line) away, leading to a reduction of light coupled into the CCW direction. 
More and more input light couples into the CW direction as the laser approaches the CW resonance, with the CCW resonance being pushed away further.
Correspondingly, the phase transmission profile of the CCW resonance is also pushed away from the input laser (as shown in the lower panel of Fig.~\ref{toyModel}(e)). However, after a certain detuning, self-phase modulation hinders the laser from approaching the CW resonance anymore, and the cross-phase modulation from the CCW direction becomes dominant, causing a reduction of the coupled power mismatch in the two directions. 
The characteristic formation of symmetry breaking "bubbles" can be seen in Fig.~\ref{toyModel}(f) and Fig.~\ref{toyModel}(i) for both power and phase profiles of the coupled fields as a function of detuning from the cold-resonances. 
Since SSB is seeded by fluctuations of the input laser, the relative dominance of the fields in the SSB region is random. Contrary to the case in Figs.~\ref{toyModel}(d-f), Fig.~\ref{toyModel}(g-i) represent the case where the CCW direction dominates.\\

\par
\begin{figure}[t]
\includegraphics[width=1\columnwidth]{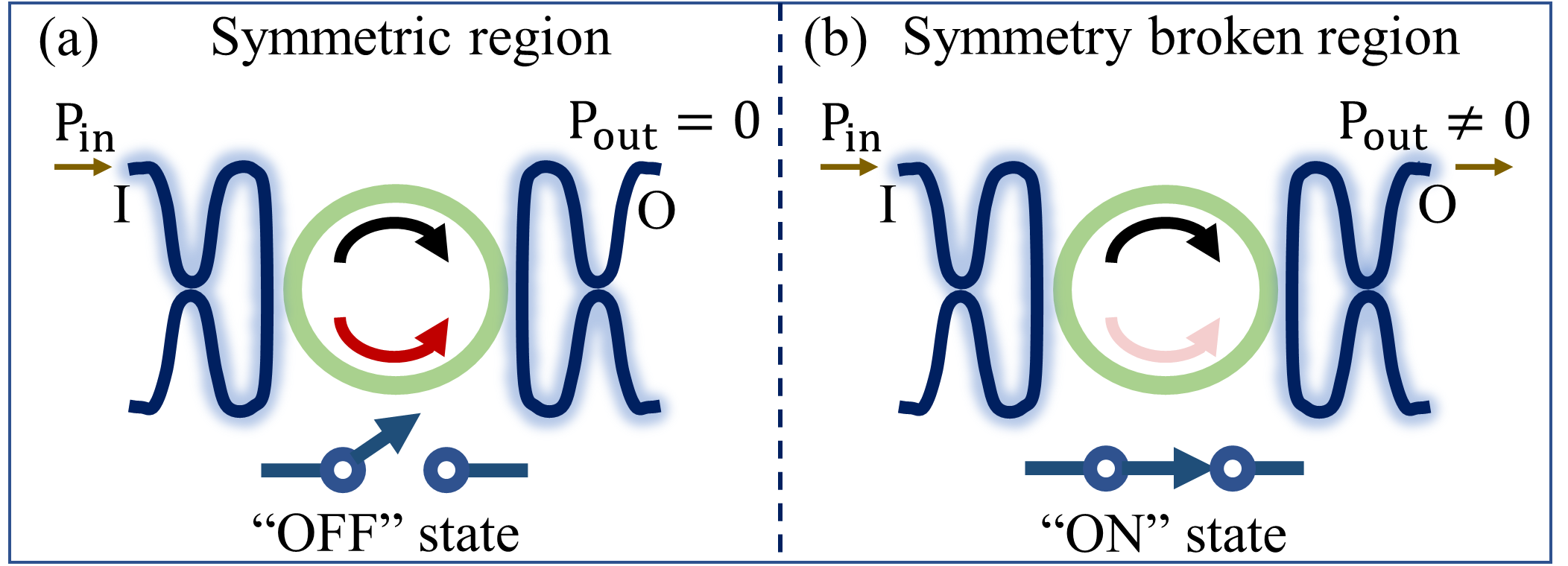}
\caption {\textit{Schematics of the optical switch.} (a) Symmetrical light intensities of the counter-propagating fields leads to constructive interference in interferometer at the output port, corresponding to the ``OFF'' state. (b) Breaking this symmetry (in this case with suppressed transmission in counterclockwise direction), prevents the destructive interference, leading to the ``ON'' state of the switch. I: input port, O: output port.}
\label{schematics}
\end{figure}

\begin{figure*}[t]
\includegraphics[width=2\columnwidth]{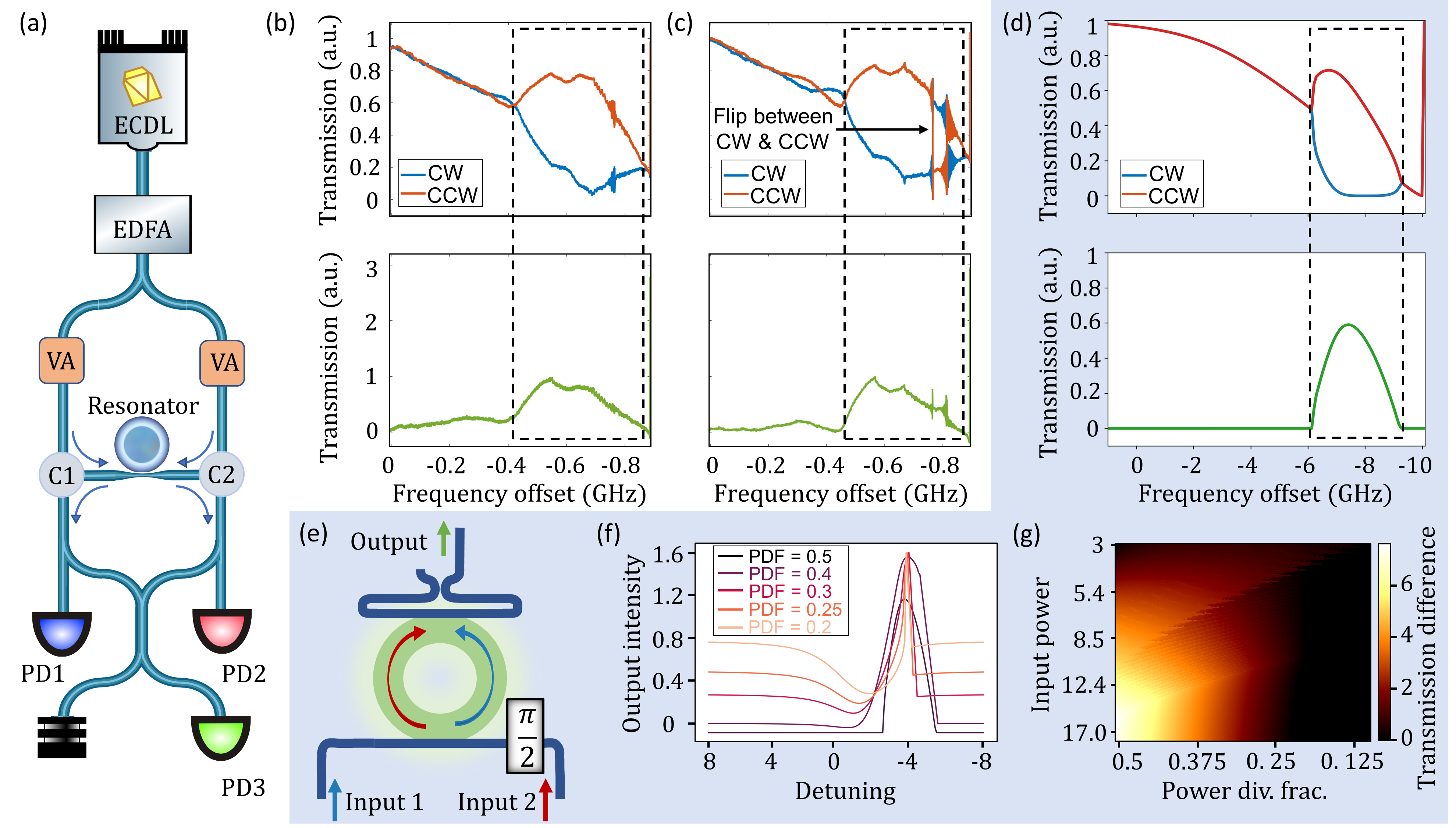}
\caption {\textit{Demonstrations of switching.} (a) Experimental setup for observing the switching effect during spontaneous symmetry breaking (SSB) in a microresonator. ECDL: External cavity diode laser, EDFA: Erbium-doped fiber amplifier, VA: Variable attenuator, C: Circulator, PD: Photodiode. 
The top panels of (b) and (c) show two experimentally observed transmission spectra of the circulating fields with their respective bottom panels showing the inteference signal from the two output directions recorded with PD3. (c) A change of dominance of the fields within the intensity-SSB bubble (marked with an arrow in the top panel) arises as the system jumps between the two stable configurations, however, this does not affect the switching mechanism. (d) Simulation of the transmitted power as a function of the laser detuning frequency is shown in the top panel, with its switching action in the bottom panel. Panel (e) depicts schematics of the switching setup used for analytical modeling. Panel (f) shows theoretical predictions of the switching effect with increasing input power imbalance. The power division factor (PDF) signifies the fraction of the total input power going to one of the input ports. Thus, the fraction of the power going to the other input port is $(1 - \text{PDF})$. (g) Varying input powers in two directions and their imbalance affect the transmission difference (defined in text) of the switch.}
\label{SwitchingExp_v2}
\end{figure*}

 \noindent \textbf{Switching effects} -- The Kerr-shifts of resonances, that cause SSB of phases in the case of identical input fields result in an enhancement of phase asymmetry between counter-propagating fields when the inputs have non-zero relative phase. This has been employed in this paper together with the SSB in intensities, to demonstrate a novel all-optical switch, as shown in Fig.~\ref{schematics}. In simple words, occurrence of SSB in intensities of the circulating fields triggers light transmission through the system, replicating the``ON'' state of a switch. Otherwise, the system does not transmit any light, corresponding to the ``OFF" state.\\
\indent The experimental setup for observing the switching effect is presented in Fig.~\ref{SwitchingExp_v2}(a). The amplified laser input at $1550$~nm is split into two halves and the evanescent fields are coupled bidirectionally into a microresonator via a tapered optical fiber. 
To access the light waves coming out of the resonator, two optical circulators are used. Ten percent of the output from each arm goes to a photodiode (PD1 or PD2) to observe the counter-propagating transmission spectrum. The rest (90\%) from both arms interferes at a 3-db-coupler, an output from which is connected to a photodiode PD3 to examine the switching effect. Here, the experiment is conducted using a $0.9$-mm-radius fused silica rod resonator with a quality factor of ${10}^8$ fabricated via $\text{CO}_2$ laser machining~\cite{del2013laser}.

\indent In the upper panel of Fig.~\ref{SwitchingExp_v2}(b), the experiments exhibit SSB between the coupled powers of the fields in the two directions below a certain detuning. The upper panel of Fig.~\ref{SwitchingExp_v2}(c) shows an SSB bubble with interchanges between the dominant and suppressed directions of circulating powers.
The lower panels of Fig.~\ref{SwitchingExp_v2}(b) and (c) show the interference of the two outputs of the two directions. Below the SSB threshold, the output is close to zero. This can be attributed to the destructive interference at the output (PD3), which comes from the two successive $3$-dB couplers (one that splits the input into two halves, one that interferes the two outputs from the resonators) each mixing the respective input fields after introducing a relative quadrature phase difference between them. Higher non-zero outputs are obtained during the period of intensity-SSB (enclosed by dashed black boxes in Fig.~\ref{SwitchingExp_v2}) as a result of a change in the relative phases of the two directions preventing a destructive interference at the output $3$-dB coupler. Therefore, light reaches PD3, only in the presence of intensity-SSB in the system, enabling the use as an indicator for symmetry breaking and optical switch. The transmission bandwidth of the switch is the same as the width of the intensity-SSB region. We term the maximum
(minimum) output at the interfering port to be the “high” (“low”) of the switch and the difference between them as the transmission difference of the switch.\\
\indent Simulated results support the experiments, as shown in Fig.~\ref{SwitchingExp_v2}(d). The upper panel shows the intensity profiles of the counter-propagating fields and the lower panel depicts the interference pattern showing the switching effect. The simplified model used for simulation is shown in Fig.~\ref{SwitchingExp_v2}(e).\\
\indent Non-uniform power splitting at the two output ports of the first 3-dB coupler, results in inequality in the input powers to the two counter-propagating directions of the resonator. In this case, for lower detuning values, the interference at the output exhibits a non-zero baseline in the absence of symmetry breaking. However, the interference output increases with increasing detuning, as seen in Fig.~\ref{SwitchingExp_v2}(f). Such a system can still be used as a switch, where the non-zero baseline can be set as the ``low" value of the switch. However, Fig.~\ref{SwitchingExp_v2}(f) depicts that with increasing imbalance, the transmission difference of the switch and the bandwidth decrease after an initial increase. Figure~\ref{SwitchingExp_v2}(g) summarises the transmission difference of the switch as a function of total input power and fraction of the total input power power pumping one of the input directions.\\\\

\par \noindent \textbf{SSB of phases} -- An initial phase difference of $\pi/2$ between the two input lasers reaching the resonator gets enhanced during the SSB of circulating intensities. A standing wave is generated in the coupling waveguide (tapered fiber) as light fields are travelling in opposite directions through it. In the presence of an anti-node of the standing wave at the coupling point, the two input fields coupled into the resonator are considered to be in phase. Under this condition, we observe SSB of the phases, which we explain analytically in this section.\\
\indent We first consider a high-$Q$ ring resonator made of a material having a $\chi^{(3)}$ nonlinearity. Input lasers with normalized amplitude $f$ are provided from both directions into the resonator at normalized detunings of $\zeta_0$ through a coupling waveguide. As a result, CW (CCW) light propagates through the resonator with normalized slowly-varying optical field amplitude $E_{+}$ ($E_{-}$). The homogeneous solution states within the resonator can be modeled using the normalized coupled Lugaito-Lefever equations (LLE)~\cite{lugiato1987spatial}:
\begin{equation}
\frac{\partial E_{\pm}}{\partial t} = - E_{\pm} - i\zeta_0 E_{\pm} + i|E_{\pm}|^2E_{\pm} + i2|E_{\mp}|^2E_{\pm} + f.
\label{LLEquations}
\end{equation}
The third and the fourth terms on the RHS describe the self- and cross-phase modulation respectively.

\begin{figure}
\includegraphics[width=0.8\columnwidth]{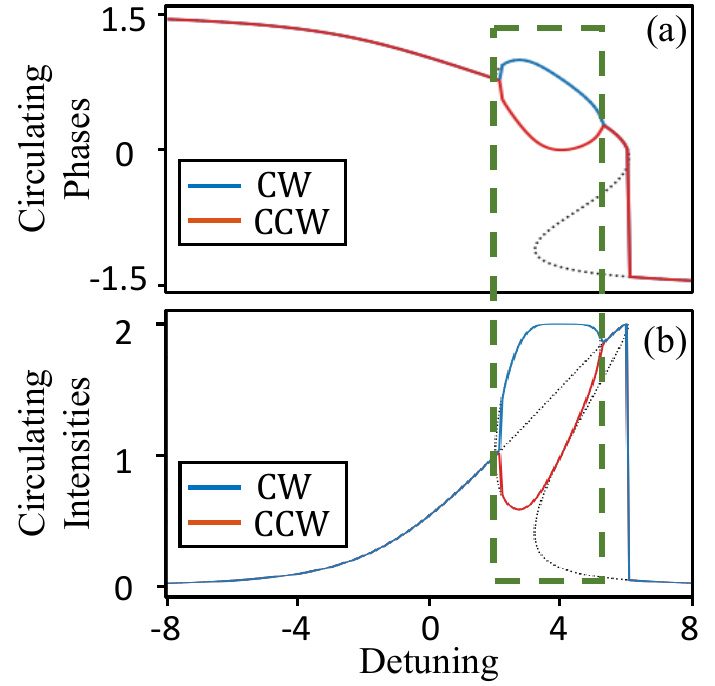}
\caption {\textit{Simulated solutions for symmetry breaking in phases.} Panels (a,b) show simulated results of circulating phases and intensities in the two directions as a function of laser detuning within a microresonator respectively, with each demonstrating spontaneous symmetry breaking (SSB). The black dotted lines show analytical solutions, the red and blue solid lines represent the circulating fields. The green dashed box signifies the concurrence of the SSBs in phases and intensities.}
\label{PhaseSSBSimsM}
\end{figure}

\indent Numerical simulations and analytical (see Appendices A \& B) solutions of Eq.~\eqref{LLEquations} demonstrate occurrences of SSB bubbles both in phases and intensities (Fig.~\ref{PhaseSSBSimsM}(a,b)) of the circulating fields for an input power higher than the SSB threshold. At higher input powers, we observe several different solutions (see Appendices B and C).
\\

\par
\begin{figure*}[t]
\includegraphics[width=2\columnwidth]{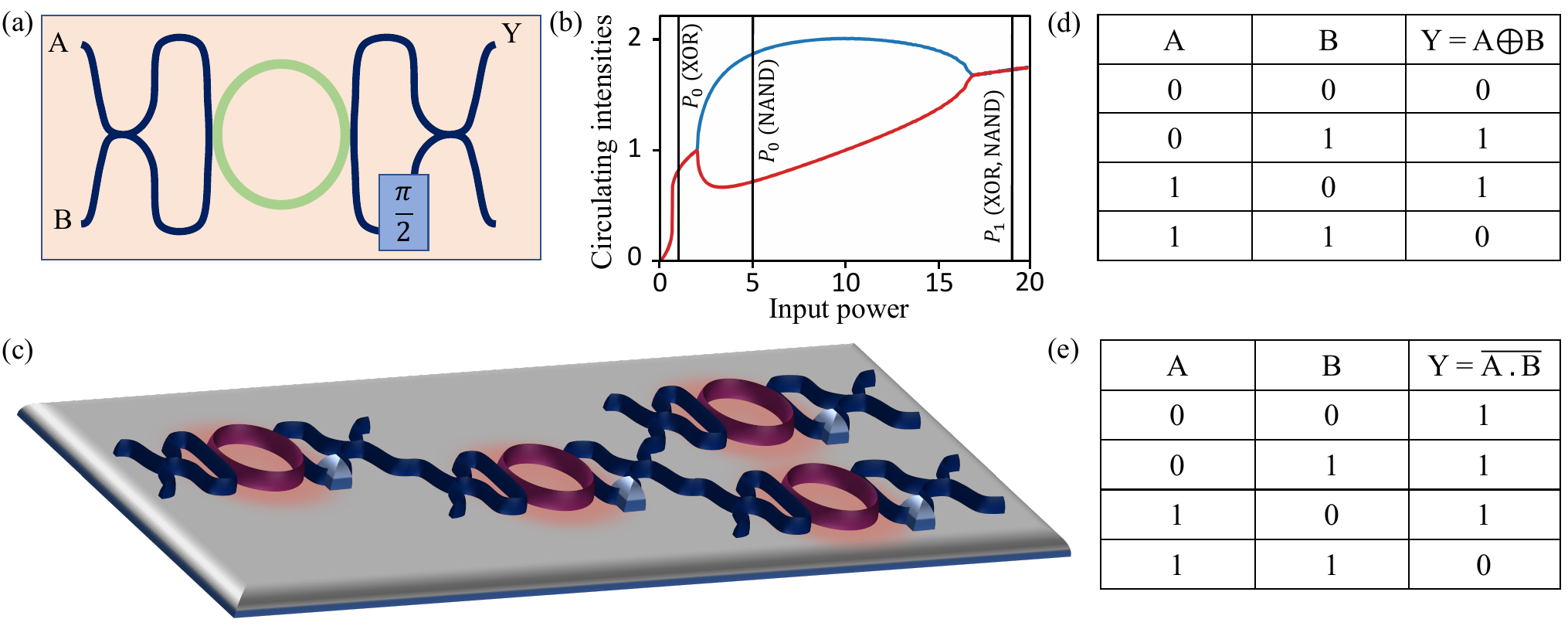}
\caption {\textit{Proposal for optical logic gates.} Panel (a) depicts schematics of the proposed optical logic gate setup. A and B represent the two inputs and Y represents the output. (b) shows an input power scan of the circulating field intensities. Different horizontal lines show different input power levels required for $1$ or $0$ logic levels for XOR or NAND gate. (c) Schematic of a photonic integrated circuit with multiple logic gates. Truth tables of XOR and NAND gates are shown in panel (d) and (e).}
\label{Logic_gates}
\end{figure*}

 \noindent \textbf{Optical logic gates} -- Here, we propose a novel structure, as shown in Fig.~\ref{Logic_gates}(a), capable of performing logical operations on integrated platforms. The operating principle is driven by the Kerr-enhancement of phase asymmetries together with SSB of intensities of the circulating fields.\\
\indent Figure~\ref{Logic_gates}(b) depicts the operating principles of XOR and NAND gates. The latter is a universal gate and can be cascaded to generate any other logic circuit. Figure~\ref{Logic_gates}(c) illustrates such a cascaded system. 
For the XOR gate, the input powers just before ($P_{0}(\text{XOR})$) and after ($P_{1}(\text{XOR})$) the SSB region are assigned as $0$ and $1$ logic inputs to the gate. An average of the input powers reaches the microresonator in each direction and whenever it lies in the intensity-SSB region (when either of the inputs is logic $1$), the transmission is non-zero, corresponding to a logic output of $1$, similar to the XOR gate truth table (Fig.~\ref{Logic_gates}(d)).\\
\indent Similarly, for the NAND gate, as seen in Fig.~\ref{Logic_gates}(b), any input power value within the SSB region ($P_{0}(\text{NAND})$) is chosen to be logic $0$, whereas an input power beyond the SSB region ($P_{1}(\text{NAND})$) is selected as logic $1$. These input power levels are chosen such that the mean power reaching the resonator lies in the SSB region when either one or both inputs correspond to a logic $0$, ($P_{0}(\text{NAND})$), resulting in non-zero transmission. This leads to the NAND gate truth table in Fig.~\ref{Logic_gates}(e). This proposed system can be easily integrated on chips to serve as a simple platform for all-optical computing. In case of excess power loss, signals could be re-amplified with integrated amplifiers~\cite{riemensberger2022photonic}.\\

\par \noindent \textbf{Discussion and outlook} -- In summary, we investigate the evolution of phases of the circulating fields in a bidirectionally pumped resonator during spontaneous symmetry breaking of intensities. These phase responses can be utlized for realizing components for photonic data processing. In a proof-of-principle we demonstrate an all-optical switch that is ``ON'' during the symmetry breaking of intensities. In addition, we propose designs for all-optical logic gates. Moreover, in Appendix~B we highlight that intensity asymmetry in a coupled resonator system~\cite{ghosh2023four} is also associated with phase asymmetry.\\
\indent Recent advances in fabrication techniques~\cite{zhang2024low, chiles2018deuterated} made it feasible to realize complex photonic devices on-chip. Phase effects in dispersion engineered integrated systems~\cite{pal2023machine, li2020real, fujii2020dispersion, bi2023chip} with symmetry broken solitons~\cite{Xu2021, Xu:22} enables the exploration of novel dynamics.
Moreover, SSB bifurcation points of the phase and intensity of the circulating fields can be utilized for microresonator based sensors~\cite{yan2024real}. Thus, the proposed devices could become promising candidates for various all-optical systems including optical neural networks, all-optical routers, and quantum information processors.\\
We would like to draw the reader’s attention to the following complementary work that investigates phase effects of SSB~\cite{anashkina2024phase}.
\par \noindent \textbf{Acknowledgements} -- This work was funded by the European Union’s H2020 ERC Starting Grant ``CounterLight” 756966, the Max Planck Society, and the Max Planck School of Photonics.

AG did the theoretical calculations with help from AP. AP fabricated the samples, performed the experiment with help from AG. All authors discussed the results. AG, AP, and PD wrote the manuscript with input from all the authors.\\\\\\


\bibliography{Refs}

\begin{thebibliography}{53}%
\makeatletter
\providecommand \@ifxundefined [1]{%
 \@ifx{#1\undefined}
}%
\providecommand \@ifnum [1]{%
 \ifnum #1\expandafter \@firstoftwo
 \else \expandafter \@secondoftwo
 \fi
}%
\providecommand \@ifx [1]{%
 \ifx #1\expandafter \@firstoftwo
 \else \expandafter \@secondoftwo
 \fi
}%
\providecommand \natexlab [1]{#1}%
\providecommand \enquote  [1]{``#1''}%
\providecommand \bibnamefont  [1]{#1}%
\providecommand \bibfnamefont [1]{#1}%
\providecommand \citenamefont [1]{#1}%
\providecommand \href@noop [0]{\@secondoftwo}%
\providecommand \href [0]{\begingroup \@sanitize@url \@href}%
\providecommand \@href[1]{\@@startlink{#1}\@@href}%
\providecommand \@@href[1]{\endgroup#1\@@endlink}%
\providecommand \@sanitize@url [0]{\catcode `\\12\catcode `\$12\catcode `\&12\catcode `\#12\catcode `\^12\catcode `\_12\catcode `\%12\relax}%
\providecommand \@@startlink[1]{}%
\providecommand \@@endlink[0]{}%
\providecommand \url  [0]{\begingroup\@sanitize@url \@url }%
\providecommand \@url [1]{\endgroup\@href {#1}{\urlprefix }}%
\providecommand \urlprefix  [0]{URL }%
\providecommand \Eprint [0]{\href }%
\providecommand \doibase [0]{http://dx.doi.org/}%
\providecommand \selectlanguage [0]{\@gobble}%
\providecommand \bibinfo  [0]{\@secondoftwo}%
\providecommand \bibfield  [0]{\@secondoftwo}%
\providecommand \translation [1]{[#1]}%
\providecommand \BibitemOpen [0]{}%
\providecommand \bibitemStop [0]{}%
\providecommand \bibitemNoStop [0]{.\EOS\space}%
\providecommand \EOS [0]{\spacefactor3000\relax}%
\providecommand \BibitemShut  [1]{\csname bibitem#1\endcsname}%
\let\auto@bib@innerbib\@empty
\bibitem [{\citenamefont {Gan} \emph {et~al.}(2015)\citenamefont {Gan}, \citenamefont {Zhao}, \citenamefont {Wang}, \citenamefont {Mao}, \citenamefont {Fang}, \citenamefont {Han}, and \citenamefont {Zhao}}]{Gan:15}%
  \BibitemOpen
  \bibfield  {author} {\bibinfo {author} {\bibfnamefont {X.}~\bibnamefont {Gan}}, \bibinfo {author} {\bibfnamefont {C.}~\bibnamefont {Zhao}}, \bibinfo {author} {\bibfnamefont {Y.}~\bibnamefont {Wang}}, \bibinfo {author} {\bibfnamefont {D.}~\bibnamefont {Mao}}, \bibinfo {author} {\bibfnamefont {L.}~\bibnamefont {Fang}}, \bibinfo {author} {\bibfnamefont {L.}~\bibnamefont {Han}},  and \bibinfo {author} {\bibfnamefont {J.}~\bibnamefont {Zhao}}, }\bibfield  {title} {\enquote {\bibinfo {title} {Graphene-assisted all-fiber phase shifter and switching},} }\href {\doibase 10.1364/OPTICA.2.000468} {\bibfield  {journal} {\bibinfo  {journal} {Optica} }\textbf {\bibinfo {volume} {2}}, \bibinfo {pages} {468--471} (\bibinfo {year} {2015})}\BibitemShut {NoStop}%
\bibitem [{\citenamefont {Clark} \emph {et~al.}(2009)\citenamefont {Clark}, \citenamefont {Fulconis}, \citenamefont {Rarity}, \citenamefont {Wadsworth}, and \citenamefont {O'Brien}}]{PhysRevA.79.030303}%
  \BibitemOpen
  \bibfield  {author} {\bibinfo {author} {\bibfnamefont {A.~S.} \bibnamefont {Clark}}, \bibinfo {author} {\bibfnamefont {J.}~\bibnamefont {Fulconis}}, \bibinfo {author} {\bibfnamefont {J.~G.} \bibnamefont {Rarity}}, \bibinfo {author} {\bibfnamefont {W.~J.} \bibnamefont {Wadsworth}},  and \bibinfo {author} {\bibfnamefont {J.~L.} \bibnamefont {O'Brien}}, }\bibfield  {title} {\enquote {\bibinfo {title} {All-optical-fiber polarization-based quantum logic gate},} }\href {\doibase 10.1103/PhysRevA.79.030303} {\bibfield  {journal} {\bibinfo  {journal} {Phys. Rev. A} }\textbf {\bibinfo {volume} {79}}, \bibinfo {pages} {030303(R)} (\bibinfo {year} {2009})}\BibitemShut {NoStop}%
\bibitem [{\citenamefont {Tanabe} \emph {et~al.}(2005)\citenamefont {Tanabe}, \citenamefont {Notomi}, \citenamefont {Mitsugi}, \citenamefont {Shinya}, and \citenamefont {Kuramochi}}]{tanabe2005fast}%
  \BibitemOpen
  \bibfield  {author} {\bibinfo {author} {\bibfnamefont {T.}~\bibnamefont {Tanabe}}, \bibinfo {author} {\bibfnamefont {M.}~\bibnamefont {Notomi}}, \bibinfo {author} {\bibfnamefont {S.}~\bibnamefont {Mitsugi}}, \bibinfo {author} {\bibfnamefont {A.}~\bibnamefont {Shinya}},  and \bibinfo {author} {\bibfnamefont {E.}~\bibnamefont {Kuramochi}}, }\bibfield  {title} {\enquote {\bibinfo {title} {Fast bistable all-optical switch and memory on a silicon photonic crystal on-chip},} }\href@noop {} {\bibfield  {journal} {\bibinfo  {journal} {Opt. Lett.} }\textbf {\bibinfo {volume} {30}}, \bibinfo {pages} {2575--2577} (\bibinfo {year} {2005})}\BibitemShut {NoStop}%
\bibitem [{\citenamefont {Kim} \emph {et~al.}(2013)\citenamefont {Kim}, \citenamefont {Bose}, \citenamefont {Shen}, \citenamefont {Solomon}, and \citenamefont {Waks}}]{kim2013quantum}%
  \BibitemOpen
  \bibfield  {author} {\bibinfo {author} {\bibfnamefont {H.}~\bibnamefont {Kim}}, \bibinfo {author} {\bibfnamefont {R.}~\bibnamefont {Bose}}, \bibinfo {author} {\bibfnamefont {T.~C.} \bibnamefont {Shen}}, \bibinfo {author} {\bibfnamefont {G.~S.} \bibnamefont {Solomon}},  and \bibinfo {author} {\bibfnamefont {E.}~\bibnamefont {Waks}}, }\bibfield  {title} {\enquote {\bibinfo {title} {A quantum logic gate between a solid-state quantum bit and a photon},} }\href@noop {} {\bibfield  {journal} {\bibinfo  {journal} {Nat. Photonics} }\textbf {\bibinfo {volume} {7}}, \bibinfo {pages} {373--377} (\bibinfo {year} {2013})}\BibitemShut {NoStop}%
\bibitem [{\citenamefont {Guo} \emph {et~al.}(2022)\citenamefont {Guo}, \citenamefont {Sekine}, \citenamefont {Ledezma}, \citenamefont {Nehra}, \citenamefont {Dean}, \citenamefont {Roy}, \citenamefont {Gray}, \citenamefont {Jahani}, and \citenamefont {Marandi}}]{guo2022femtojoule}%
  \BibitemOpen
  \bibfield  {author} {\bibinfo {author} {\bibfnamefont {Q.}~\bibnamefont {Guo}}, \bibinfo {author} {\bibfnamefont {R.}~\bibnamefont {Sekine}}, \bibinfo {author} {\bibfnamefont {L.}~\bibnamefont {Ledezma}}, \bibinfo {author} {\bibfnamefont {R.}~\bibnamefont {Nehra}}, \bibinfo {author} {\bibfnamefont {D.~J.} \bibnamefont {Dean}}, \bibinfo {author} {\bibfnamefont {A.}~\bibnamefont {Roy}}, \bibinfo {author} {\bibfnamefont {R.~M.} \bibnamefont {Gray}}, \bibinfo {author} {\bibfnamefont {S.}~\bibnamefont {Jahani}},  and \bibinfo {author} {\bibfnamefont {A.}~\bibnamefont {Marandi}}, }\bibfield  {title} {\enquote {\bibinfo {title} {Femtojoule femtosecond all-optical switching in lithium niobate nanophotonics},} }\href@noop {} {\bibfield  {journal} {\bibinfo  {journal} {Nat. Photonics} }\textbf {\bibinfo {volume} {16}}, \bibinfo {pages} {625--631} (\bibinfo {year} {2022})}\BibitemShut {NoStop}%
\bibitem [{\citenamefont {Ono} \emph {et~al.}(2020)\citenamefont {Ono}, \citenamefont {Hata}, \citenamefont {Tsunekawa}, \citenamefont {Nozaki}, \citenamefont {Sumikura}, \citenamefont {Chiba}, and \citenamefont {Notomi}}]{ono2020ultrafast}%
  \BibitemOpen
  \bibfield  {author} {\bibinfo {author} {\bibfnamefont {M.}~\bibnamefont {Ono}}, \bibinfo {author} {\bibfnamefont {M.}~\bibnamefont {Hata}}, \bibinfo {author} {\bibfnamefont {M.}~\bibnamefont {Tsunekawa}}, \bibinfo {author} {\bibfnamefont {K.}~\bibnamefont {Nozaki}}, \bibinfo {author} {\bibfnamefont {H.}~\bibnamefont {Sumikura}}, \bibinfo {author} {\bibfnamefont {H.}~\bibnamefont {Chiba}},  and \bibinfo {author} {\bibfnamefont {M.}~\bibnamefont {Notomi}}, }\bibfield  {title} {\enquote {\bibinfo {title} {Ultrafast and energy-efficient all-optical switching with graphene-loaded deep-subwavelength plasmonic waveguides},} }\href@noop {} {\bibfield  {journal} {\bibinfo  {journal} {Nat. Photonics} }\textbf {\bibinfo {volume} {14}}, \bibinfo {pages} {37--43} (\bibinfo {year} {2020})}\BibitemShut {NoStop}%
\bibitem [{\citenamefont {Hoessbacher} \emph {et~al.}(2014)\citenamefont {Hoessbacher}, \citenamefont {Fedoryshyn}, \citenamefont {Emboras}, \citenamefont {Melikyan}, \citenamefont {Kohl}, \citenamefont {Hillerkuss}, \citenamefont {Hafner}, and \citenamefont {Leuthold}}]{Hoessbacher:14}%
  \BibitemOpen
  \bibfield  {author} {\bibinfo {author} {\bibfnamefont {C.}~\bibnamefont {Hoessbacher}}, \bibinfo {author} {\bibfnamefont {Y.}~\bibnamefont {Fedoryshyn}}, \bibinfo {author} {\bibfnamefont {A.}~\bibnamefont {Emboras}}, \bibinfo {author} {\bibfnamefont {A.}~\bibnamefont {Melikyan}}, \bibinfo {author} {\bibfnamefont {M.}~\bibnamefont {Kohl}}, \bibinfo {author} {\bibfnamefont {D.}~\bibnamefont {Hillerkuss}}, \bibinfo {author} {\bibfnamefont {C.}~\bibnamefont {Hafner}},  and \bibinfo {author} {\bibfnamefont {J.}~\bibnamefont {Leuthold}}, }\bibfield  {title} {\enquote {\bibinfo {title} {The plasmonic memristor: a latching optical switch},} }\href {\doibase 10.1364/OPTICA.1.000198} {\bibfield  {journal} {\bibinfo  {journal} {Optica} }\textbf {\bibinfo {volume} {1}}, \bibinfo {pages} {198--202} (\bibinfo {year} {2014})}\BibitemShut {NoStop}%
\bibitem [{\citenamefont {Wei} \emph {et~al.}(2011)\citenamefont {Wei}, \citenamefont {Wang}, \citenamefont {Tian}, \citenamefont {K{\"a}ll}, and \citenamefont {Xu}}]{wei2011cascaded}%
  \BibitemOpen
  \bibfield  {author} {\bibinfo {author} {\bibfnamefont {H.}~\bibnamefont {Wei}}, \bibinfo {author} {\bibfnamefont {Z.}~\bibnamefont {Wang}}, \bibinfo {author} {\bibfnamefont {X.}~\bibnamefont {Tian}}, \bibinfo {author} {\bibfnamefont {M.}~\bibnamefont {K{\"a}ll}},  and \bibinfo {author} {\bibfnamefont {H.}~\bibnamefont {Xu}}, }\bibfield  {title} {\enquote {\bibinfo {title} {Cascaded logic gates in nanophotonic plasmon networks},} }\href@noop {} {\bibfield  {journal} {\bibinfo  {journal} {Nat. Commun.} }\textbf {\bibinfo {volume} {2}}, \bibinfo {pages} {387} (\bibinfo {year} {2011})}\BibitemShut {NoStop}%
\bibitem [{\citenamefont {Almeida} \emph {et~al.}(2004)\citenamefont {Almeida}, \citenamefont {Barrios}, \citenamefont {Panepucci}, and \citenamefont {Lipson}}]{almeida2004all}%
  \BibitemOpen
  \bibfield  {author} {\bibinfo {author} {\bibfnamefont {V.~R.} \bibnamefont {Almeida}}, \bibinfo {author} {\bibfnamefont {C.~A.} \bibnamefont {Barrios}}, \bibinfo {author} {\bibfnamefont {R.~R.} \bibnamefont {Panepucci}},  and \bibinfo {author} {\bibfnamefont {M.}~\bibnamefont {Lipson}}, }\bibfield  {title} {\enquote {\bibinfo {title} {All-optical control of light on a silicon chip},} }\href@noop {} {\bibfield  {journal} {\bibinfo  {journal} {Nature} }\textbf {\bibinfo {volume} {431}}, \bibinfo {pages} {1081--1084} (\bibinfo {year} {2004})}\BibitemShut {NoStop}%
\bibitem [{\citenamefont {Raja} \emph {et~al.}(2021)\citenamefont {Raja}, \citenamefont {Lange}, \citenamefont {Karpov}, \citenamefont {Shi}, \citenamefont {Fu}, \citenamefont {Behrendt}, \citenamefont {Cletheroe}, \citenamefont {Lukashchuk}, \citenamefont {Haller}, \citenamefont {Karinou} \emph {et~al.}}]{raja2021ultrafast}%
  \BibitemOpen
  \bibfield  {author} {\bibinfo {author} {\bibfnamefont {A.~S.} \bibnamefont {Raja}}, \bibinfo {author} {\bibfnamefont {S.}~\bibnamefont {Lange}}, \bibinfo {author} {\bibfnamefont {M.}~\bibnamefont {Karpov}}, \bibinfo {author} {\bibfnamefont {K.}~\bibnamefont {Shi}}, \bibinfo {author} {\bibfnamefont {X.}~\bibnamefont {Fu}}, \bibinfo {author} {\bibfnamefont {R.}~\bibnamefont {Behrendt}}, \bibinfo {author} {\bibfnamefont {D.}~\bibnamefont {Cletheroe}}, \bibinfo {author} {\bibfnamefont {A.}~\bibnamefont {Lukashchuk}}, \bibinfo {author} {\bibfnamefont {I.}~\bibnamefont {Haller}}, \bibinfo {author} {\bibfnamefont {F.}~\bibnamefont {Karinou}},  \emph {et~al.}, }\bibfield  {title} {\enquote {\bibinfo {title} {Ultrafast optical circuit switching for data centers using integrated soliton microcombs},} }\href@noop {} {\bibfield  {journal} {\bibinfo  {journal} {Nat. Commun.} }\textbf {\bibinfo {volume} {12}}, \bibinfo {pages} {5867} (\bibinfo {year} {2021})}\BibitemShut {NoStop}%
\bibitem [{\citenamefont {Xu} and \citenamefont {Soref}(2011)}]{Xu:11}%
  \BibitemOpen
  \bibfield  {author} {\bibinfo {author} {\bibfnamefont {Q.}~\bibnamefont {Xu}} and \bibinfo {author} {\bibfnamefont {R.}~\bibnamefont {Soref}}, }\bibfield  {title} {\enquote {\bibinfo {title} {Reconfigurable optical directed-logic circuits using microresonator-based optical switches},} }\href {\doibase 10.1364/OE.19.005244} {\bibfield  {journal} {\bibinfo  {journal} {Opt. Express} }\textbf {\bibinfo {volume} {19}}, \bibinfo {pages} {5244--5259} (\bibinfo {year} {2011})}\BibitemShut {NoStop}%
\bibitem [{\citenamefont {Van} \emph {et~al.}(2002)\citenamefont {Van}, \citenamefont {Ibrahim}, \citenamefont {Ritter}, \citenamefont {Absil}, \citenamefont {Johnson}, \citenamefont {Grover}, \citenamefont {Goldhar}, and \citenamefont {Ho}}]{van2002all}%
  \BibitemOpen
  \bibfield  {author} {\bibinfo {author} {\bibfnamefont {V.}~\bibnamefont {Van}}, \bibinfo {author} {\bibfnamefont {T.}~\bibnamefont {Ibrahim}}, \bibinfo {author} {\bibfnamefont {K.}~\bibnamefont {Ritter}}, \bibinfo {author} {\bibfnamefont {P.}~\bibnamefont {Absil}}, \bibinfo {author} {\bibfnamefont {F.}~\bibnamefont {Johnson}}, \bibinfo {author} {\bibfnamefont {R.}~\bibnamefont {Grover}}, \bibinfo {author} {\bibfnamefont {J.}~\bibnamefont {Goldhar}},  and \bibinfo {author} {\bibfnamefont {P.-T.} \bibnamefont {Ho}}, }\bibfield  {title} {\enquote {\bibinfo {title} {All-optical nonlinear switching in {G}a{A}s-{A}l{G}a{A}s microring resonators},} }\href@noop {} {\bibfield  {journal} {\bibinfo  {journal} {IEEE Photonics Technol. Lett.} }\textbf {\bibinfo {volume} {14}}, \bibinfo {pages} {74--76} (\bibinfo {year} {2002})}\BibitemShut {NoStop}%
\bibitem [{\citenamefont {Bino} \emph {et~al.}(2018)\citenamefont {Bino}, \citenamefont {Silver}, \citenamefont {Woodley}, \citenamefont {Stebbings}, \citenamefont {Zhao}, and \citenamefont {Del'Haye}}]{DelBino:18}%
  \BibitemOpen
  \bibfield  {author} {\bibinfo {author} {\bibfnamefont {L.~D.} \bibnamefont {Bino}}, \bibinfo {author} {\bibfnamefont {J.~M.} \bibnamefont {Silver}}, \bibinfo {author} {\bibfnamefont {M.~T.~M.} \bibnamefont {Woodley}}, \bibinfo {author} {\bibfnamefont {S.~L.} \bibnamefont {Stebbings}}, \bibinfo {author} {\bibfnamefont {X.}~\bibnamefont {Zhao}},  and \bibinfo {author} {\bibfnamefont {P.}~\bibnamefont {Del'Haye}}, }\bibfield  {title} {\enquote {\bibinfo {title} {Microresonator isolators and circulators based on the intrinsic nonreciprocity of the {K}err effect},} }\href {\doibase 10.1364/OPTICA.5.000279} {\bibfield  {journal} {\bibinfo  {journal} {Optica} }\textbf {\bibinfo {volume} {5}}, \bibinfo {pages} {279--282} (\bibinfo {year} {2018})}\BibitemShut {NoStop}%
\bibitem [{\citenamefont {Godbole} \emph {et~al.}(2016)\citenamefont {Godbole}, \citenamefont {Dali}, \citenamefont {Janyani}, \citenamefont {Tanabe}, and \citenamefont {Singh}}]{godbole2016all}%
  \BibitemOpen
  \bibfield  {author} {\bibinfo {author} {\bibfnamefont {A.}~\bibnamefont {Godbole}}, \bibinfo {author} {\bibfnamefont {P.~P.} \bibnamefont {Dali}}, \bibinfo {author} {\bibfnamefont {V.}~\bibnamefont {Janyani}}, \bibinfo {author} {\bibfnamefont {T.}~\bibnamefont {Tanabe}},  and \bibinfo {author} {\bibfnamefont {G.}~\bibnamefont {Singh}}, }\bibfield  {title} {\enquote {\bibinfo {title} {All optical scalable logic gates using $\text{{S}i}_3\text{{N}}_4$ microring resonators},} }\href@noop {} {\bibfield  {journal} {\bibinfo  {journal} {IEEE J. Sel. Top. Quantum Electron.} }\textbf {\bibinfo {volume} {22}}, \bibinfo {pages} {326--333} (\bibinfo {year} {2016})}\BibitemShut {NoStop}%
\bibitem [{\citenamefont {Sethi} and \citenamefont {Roy}(2014)}]{sethi2014all}%
  \BibitemOpen
  \bibfield  {author} {\bibinfo {author} {\bibfnamefont {P.}~\bibnamefont {Sethi}} and \bibinfo {author} {\bibfnamefont {S.}~\bibnamefont {Roy}}, }\bibfield  {title} {\enquote {\bibinfo {title} {All-optical ultrafast {XOR/XNOR} logic gates, binary counter, and double-bit comparator with silicon microring resonators},} }\href@noop {} {\bibfield  {journal} {\bibinfo  {journal} {Appl. Opt.} }\textbf {\bibinfo {volume} {53}}, \bibinfo {pages} {6527--6536} (\bibinfo {year} {2014})}\BibitemShut {NoStop}%
\bibitem [{\citenamefont {Qiu} \emph {et~al.}(2020)\citenamefont {Qiu}, \citenamefont {Zhang}, \citenamefont {Zeng}, and \citenamefont {Guo}}]{qiu2020high}%
  \BibitemOpen
  \bibfield  {author} {\bibinfo {author} {\bibfnamefont {C.}~\bibnamefont {Qiu}}, \bibinfo {author} {\bibfnamefont {C.}~\bibnamefont {Zhang}}, \bibinfo {author} {\bibfnamefont {H.}~\bibnamefont {Zeng}},  and \bibinfo {author} {\bibfnamefont {T.}~\bibnamefont {Guo}}, }\bibfield  {title} {\enquote {\bibinfo {title} {High-performance graphene-on-silicon nitride all-optical switch based on a {M}ach--{Z}ehnder interferometer},} }\href@noop {} {\bibfield  {journal} {\bibinfo  {journal} {J. Light. Technol.} }\textbf {\bibinfo {volume} {39}}, \bibinfo {pages} {2099--2105} (\bibinfo {year} {2020})}\BibitemShut {NoStop}%
\bibitem [{\citenamefont {Hill} \emph {et~al.}(2004)\citenamefont {Hill}, \citenamefont {Dorren}, \citenamefont {De~Vries}, \citenamefont {Leijtens}, \citenamefont {Den~Besten}, \citenamefont {Smalbrugge}, \citenamefont {Oei}, \citenamefont {Binsma}, \citenamefont {Khoe}, and \citenamefont {Smit}}]{hill2004fast}%
  \BibitemOpen
  \bibfield  {author} {\bibinfo {author} {\bibfnamefont {M.~T.} \bibnamefont {Hill}}, \bibinfo {author} {\bibfnamefont {H.~J.} \bibnamefont {Dorren}}, \bibinfo {author} {\bibfnamefont {T.}~\bibnamefont {De~Vries}}, \bibinfo {author} {\bibfnamefont {X.~J.} \bibnamefont {Leijtens}}, \bibinfo {author} {\bibfnamefont {J.~H.} \bibnamefont {Den~Besten}}, \bibinfo {author} {\bibfnamefont {B.}~\bibnamefont {Smalbrugge}}, \bibinfo {author} {\bibfnamefont {Y.-S.} \bibnamefont {Oei}}, \bibinfo {author} {\bibfnamefont {H.}~\bibnamefont {Binsma}}, \bibinfo {author} {\bibfnamefont {G.-D.} \bibnamefont {Khoe}},  and \bibinfo {author} {\bibfnamefont {M.~K.} \bibnamefont {Smit}}, }\bibfield  {title} {\enquote {\bibinfo {title} {A fast low-power optical memory based on coupled micro-ring lasers},} }\href@noop {} {\bibfield  {journal} {\bibinfo  {journal} {Nature} }\textbf {\bibinfo {volume} {432}}, \bibinfo {pages} {206--209} (\bibinfo {year} {2004})}\BibitemShut {NoStop}%
\bibitem [{\citenamefont {Maes} \emph {et~al.}(2006)\citenamefont {Maes}, \citenamefont {Solja{\v{c}}i{\'c}}, \citenamefont {Joannopoulos}, \citenamefont {Bienstman}, \citenamefont {Baets}, \citenamefont {Gorza}, and \citenamefont {Haelterman}}]{maes2006switching}%
  \BibitemOpen
  \bibfield  {author} {\bibinfo {author} {\bibfnamefont {B.}~\bibnamefont {Maes}}, \bibinfo {author} {\bibfnamefont {M.}~\bibnamefont {Solja{\v{c}}i{\'c}}}, \bibinfo {author} {\bibfnamefont {J.~D.} \bibnamefont {Joannopoulos}}, \bibinfo {author} {\bibfnamefont {P.}~\bibnamefont {Bienstman}}, \bibinfo {author} {\bibfnamefont {R.}~\bibnamefont {Baets}}, \bibinfo {author} {\bibfnamefont {S.-P.} \bibnamefont {Gorza}},  and \bibinfo {author} {\bibfnamefont {M.}~\bibnamefont {Haelterman}}, }\bibfield  {title} {\enquote {\bibinfo {title} {Switching through symmetry breaking in coupled nonlinear micro-cavities},} }\href@noop {} {\bibfield  {journal} {\bibinfo  {journal} {Opt. Express} }\textbf {\bibinfo {volume} {14}}, \bibinfo {pages} {10678--10683} (\bibinfo {year} {2006})}\BibitemShut {NoStop}%
\bibitem [{\citenamefont {Hamel} \emph {et~al.}(2015)\citenamefont {Hamel}, \citenamefont {Haddadi}, \citenamefont {Raineri}, \citenamefont {Monnier}, \citenamefont {Beaudoin}, \citenamefont {Sagnes}, \citenamefont {Levenson}, and \citenamefont {Yacomotti}}]{hamel2015spontaneous}%
  \BibitemOpen
  \bibfield  {author} {\bibinfo {author} {\bibfnamefont {P.}~\bibnamefont {Hamel}}, \bibinfo {author} {\bibfnamefont {S.}~\bibnamefont {Haddadi}}, \bibinfo {author} {\bibfnamefont {F.}~\bibnamefont {Raineri}}, \bibinfo {author} {\bibfnamefont {P.}~\bibnamefont {Monnier}}, \bibinfo {author} {\bibfnamefont {G.}~\bibnamefont {Beaudoin}}, \bibinfo {author} {\bibfnamefont {I.}~\bibnamefont {Sagnes}}, \bibinfo {author} {\bibfnamefont {A.}~\bibnamefont {Levenson}},  and \bibinfo {author} {\bibfnamefont {A.~M.} \bibnamefont {Yacomotti}}, }\bibfield  {title} {\enquote {\bibinfo {title} {Spontaneous mirror-symmetry breaking in coupled photonic-crystal nanolasers},} }\href@noop {} {\bibfield  {journal} {\bibinfo  {journal} {Nat. Photonics} }\textbf {\bibinfo {volume} {9}}, \bibinfo {pages} {311--315} (\bibinfo {year} {2015})}\BibitemShut {NoStop}%
\bibitem [{\citenamefont {Kaplan} and \citenamefont {Meystre}(1982)}]{KAPLAN1982229}%
  \BibitemOpen
  \bibfield  {author} {\bibinfo {author} {\bibfnamefont {A.}~\bibnamefont {Kaplan}} and \bibinfo {author} {\bibfnamefont {P.}~\bibnamefont {Meystre}}, }\bibfield  {title} {\enquote {\bibinfo {title} {Directionally asymmetrical bistability in a symmetrically pumped nonlinear ring interferometer},} }\href {\doibase https://doi.org/10.1016/0030-4018(82)90267-X} {\bibfield  {journal} {\bibinfo  {journal} {Opt. Commun.} }\textbf {\bibinfo {volume} {40}}, \bibinfo {pages} {229--232} (\bibinfo {year} {1982})}\BibitemShut {NoStop}%
\bibitem [{\citenamefont {Woodley} \emph {et~al.}(2018)\citenamefont {Woodley}, \citenamefont {Silver}, \citenamefont {Hill}, \citenamefont {Copie}, \citenamefont {Del~Bino}, \citenamefont {Zhang}, \citenamefont {Oppo}, and \citenamefont {Del'Haye}}]{woodley2018universal}%
  \BibitemOpen
  \bibfield  {author} {\bibinfo {author} {\bibfnamefont {M.~T.~M.} \bibnamefont {Woodley}}, \bibinfo {author} {\bibfnamefont {J.~M.} \bibnamefont {Silver}}, \bibinfo {author} {\bibfnamefont {L.}~\bibnamefont {Hill}}, \bibinfo {author} {\bibfnamefont {F.}~\bibnamefont {Copie}}, \bibinfo {author} {\bibfnamefont {L.}~\bibnamefont {Del~Bino}}, \bibinfo {author} {\bibfnamefont {S.}~\bibnamefont {Zhang}}, \bibinfo {author} {\bibfnamefont {G.-L.} \bibnamefont {Oppo}},  and \bibinfo {author} {\bibfnamefont {P.}~\bibnamefont {Del'Haye}}, }\bibfield  {title} {\enquote {\bibinfo {title} {Universal symmetry-breaking dynamics for the {K}err interaction of counterpropagating light in dielectric ring resonators},} }\href {\doibase 10.1103/PhysRevA.98.053863} {\bibfield  {journal} {\bibinfo  {journal} {Phys. Rev. A} }\textbf {\bibinfo {volume} {98}}, \bibinfo {pages} {053863} (\bibinfo {year} {2018})}\BibitemShut {NoStop}%
\bibitem [{\citenamefont {Del~Bino} \emph {et~al.}(2017)\citenamefont {Del~Bino}, \citenamefont {Silver}, \citenamefont {Stebbings}, and \citenamefont {Del'Haye}}]{DelBino2017}%
  \BibitemOpen
  \bibfield  {author} {\bibinfo {author} {\bibfnamefont {L.}~\bibnamefont {Del~Bino}}, \bibinfo {author} {\bibfnamefont {J.~M.} \bibnamefont {Silver}}, \bibinfo {author} {\bibfnamefont {S.~L.} \bibnamefont {Stebbings}},  and \bibinfo {author} {\bibfnamefont {P.}~\bibnamefont {Del'Haye}}, }\bibfield  {title} {\enquote {\bibinfo {title} {Symmetry breaking of counter-propagating light in a nonlinear resonator},} }\href {\doibase 10.1038/srep43142} {\bibfield  {journal} {\bibinfo  {journal} {Sci. Rep.} }\textbf {\bibinfo {volume} {7}}, \bibinfo {pages} {43142} (\bibinfo {year} {2017})}\BibitemShut {NoStop}%
\bibitem [{\citenamefont {Copie} \emph {et~al.}(2019)\citenamefont {Copie}, \citenamefont {Woodley}, \citenamefont {Del~Bino}, \citenamefont {Silver}, \citenamefont {Zhang}, and \citenamefont {Del’Haye}}]{PhysRevLett.122.013905}%
  \BibitemOpen
  \bibfield  {author} {\bibinfo {author} {\bibfnamefont {F.}~\bibnamefont {Copie}}, \bibinfo {author} {\bibfnamefont {M.~T.~M.} \bibnamefont {Woodley}}, \bibinfo {author} {\bibfnamefont {L.}~\bibnamefont {Del~Bino}}, \bibinfo {author} {\bibfnamefont {J.~M.} \bibnamefont {Silver}}, \bibinfo {author} {\bibfnamefont {S.}~\bibnamefont {Zhang}},  and \bibinfo {author} {\bibfnamefont {P.}~\bibnamefont {Del’Haye}}, }\bibfield  {title} {\enquote {\bibinfo {title} {Interplay of polarization and time-reversal symmetry breaking in synchronously pumped ring resonators},} }\href {\doibase 10.1103/PhysRevLett.122.013905} {\bibfield  {journal} {\bibinfo  {journal} {Phys. Rev. Lett.} }\textbf {\bibinfo {volume} {122}}, \bibinfo {pages} {013905} (\bibinfo {year} {2019})}\BibitemShut {NoStop}%
\bibitem [{\citenamefont {Garbin} \emph {et~al.}(2020)\citenamefont {Garbin}, \citenamefont {Fatome}, \citenamefont {Oppo}, \citenamefont {Erkintalo}, \citenamefont {Murdoch}, and \citenamefont {Coen}}]{garbin2020asymmetric}%
  \BibitemOpen
  \bibfield  {author} {\bibinfo {author} {\bibfnamefont {B.}~\bibnamefont {Garbin}}, \bibinfo {author} {\bibfnamefont {J.}~\bibnamefont {Fatome}}, \bibinfo {author} {\bibfnamefont {G.-L.} \bibnamefont {Oppo}}, \bibinfo {author} {\bibfnamefont {M.}~\bibnamefont {Erkintalo}}, \bibinfo {author} {\bibfnamefont {S.~G.} \bibnamefont {Murdoch}},  and \bibinfo {author} {\bibfnamefont {S.}~\bibnamefont {Coen}}, }\bibfield  {title} {\enquote {\bibinfo {title} {Asymmetric balance in symmetry breaking},} }\href@noop {} {\bibfield  {journal} {\bibinfo  {journal} {Phys. Rev. Res.} }\textbf {\bibinfo {volume} {2}}, \bibinfo {pages} {023244} (\bibinfo {year} {2020})}\BibitemShut {NoStop}%
\bibitem [{\citenamefont {Hill} \emph {et~al.}(2024)\citenamefont {Hill}, \citenamefont {Hirmer}, \citenamefont {Campbell}, \citenamefont {Bi}, \citenamefont {Ghosh}, \citenamefont {Del’Haye}, and \citenamefont {Oppo}}]{hill2024symmetry}%
  \BibitemOpen
  \bibfield  {author} {\bibinfo {author} {\bibfnamefont {L.}~\bibnamefont {Hill}}, \bibinfo {author} {\bibfnamefont {E.-M.} \bibnamefont {Hirmer}}, \bibinfo {author} {\bibfnamefont {G.}~\bibnamefont {Campbell}}, \bibinfo {author} {\bibfnamefont {T.}~\bibnamefont {Bi}}, \bibinfo {author} {\bibfnamefont {A.}~\bibnamefont {Ghosh}}, \bibinfo {author} {\bibfnamefont {P.}~\bibnamefont {Del’Haye}},  and \bibinfo {author} {\bibfnamefont {G.-L.} \bibnamefont {Oppo}}, }\bibfield  {title} {\enquote {\bibinfo {title} {Symmetry broken vectorial {K}err frequency combs from {F}abry-{P}{\'e}rot resonators},} }\href@noop {} {\bibfield  {journal} {\bibinfo  {journal} {Commun. Phys.} }\textbf {\bibinfo {volume} {7}}, \bibinfo {pages} {82} (\bibinfo {year} {2024})}\BibitemShut {NoStop}%
\bibitem [{\citenamefont {Campbell} \emph {et~al.}(2024)\citenamefont {Campbell}, \citenamefont {Hill}, \citenamefont {Del'Haye}, and \citenamefont {Oppo}}]{campbell2024frequency}%
  \BibitemOpen
  \bibfield  {author} {\bibinfo {author} {\bibfnamefont {G.~N.} \bibnamefont {Campbell}}, \bibinfo {author} {\bibfnamefont {L.}~\bibnamefont {Hill}}, \bibinfo {author} {\bibfnamefont {P.}~\bibnamefont {Del'Haye}},  and \bibinfo {author} {\bibfnamefont {G.-L.} \bibnamefont {Oppo}}, }\bibfield  {title} {\enquote {\bibinfo {title} {Frequency comb enhancement via the self-crystallization of vectorial cavity solitons},} }\href@noop {} {\bibfield  {journal} {\bibinfo  {journal} {arXiv preprint arXiv:2403.16547} } (\bibinfo {year} {2024})}\BibitemShut {NoStop}%
\bibitem [{\citenamefont {Bino} \emph {et~al.}(2021)\citenamefont {Bino}, \citenamefont {Moroney}, and \citenamefont {Del'Haye}}]{DelBino:21}%
  \BibitemOpen
  \bibfield  {author} {\bibinfo {author} {\bibfnamefont {L.~D.} \bibnamefont {Bino}}, \bibinfo {author} {\bibfnamefont {N.}~\bibnamefont {Moroney}},  and \bibinfo {author} {\bibfnamefont {P.}~\bibnamefont {Del'Haye}}, }\bibfield  {title} {\enquote {\bibinfo {title} {Optical memories and switching dynamics of counterpropagating light states in microresonators},} }\href {\doibase 10.1364/OE.417951} {\bibfield  {journal} {\bibinfo  {journal} {Opt. Express} }\textbf {\bibinfo {volume} {29}}, \bibinfo {pages} {2193--2203} (\bibinfo {year} {2021})}\BibitemShut {NoStop}%
\bibitem [{\citenamefont {Silver} \emph {et~al.}(2021)\citenamefont {Silver}, \citenamefont {Bino}, \citenamefont {Woodley}, \citenamefont {Ghalanos}, \citenamefont {Svela}, \citenamefont {Moroney}, \citenamefont {Zhang}, \citenamefont {Grattan}, and \citenamefont {Del'Haye}}]{Silver:21}%
  \BibitemOpen
  \bibfield  {author} {\bibinfo {author} {\bibfnamefont {J.~M.} \bibnamefont {Silver}}, \bibinfo {author} {\bibfnamefont {L.~D.} \bibnamefont {Bino}}, \bibinfo {author} {\bibfnamefont {M.~T.~M.} \bibnamefont {Woodley}}, \bibinfo {author} {\bibfnamefont {G.~N.} \bibnamefont {Ghalanos}}, \bibinfo {author} {\bibfnamefont {A.~{\O}.} \bibnamefont {Svela}}, \bibinfo {author} {\bibfnamefont {N.}~\bibnamefont {Moroney}}, \bibinfo {author} {\bibfnamefont {S.}~\bibnamefont {Zhang}}, \bibinfo {author} {\bibfnamefont {K.~T.~V.} \bibnamefont {Grattan}},  and \bibinfo {author} {\bibfnamefont {P.}~\bibnamefont {Del'Haye}}, }\bibfield  {title} {\enquote {\bibinfo {title} {Nonlinear enhanced microresonator gyroscope},} }\href {\doibase 10.1364/OPTICA.426018} {\bibfield  {journal} {\bibinfo  {journal} {Optica} }\textbf {\bibinfo {volume} {8}}, \bibinfo {pages} {1219--1226} (\bibinfo {year} {2021})}\BibitemShut {NoStop}%
\bibitem [{\citenamefont {Moroney} \emph {et~al.}(2020)\citenamefont {Moroney}, \citenamefont {Bino}, \citenamefont {Woodley}, \citenamefont {Ghalanos}, \citenamefont {Silver}, \citenamefont {Svela}, \citenamefont {Zhang}, and \citenamefont {Del'Haye}}]{Moroney:20}%
  \BibitemOpen
  \bibfield  {author} {\bibinfo {author} {\bibfnamefont {N.}~\bibnamefont {Moroney}}, \bibinfo {author} {\bibfnamefont {L.~D.} \bibnamefont {Bino}}, \bibinfo {author} {\bibfnamefont {M.~T.~M.} \bibnamefont {Woodley}}, \bibinfo {author} {\bibfnamefont {G.~N.} \bibnamefont {Ghalanos}}, \bibinfo {author} {\bibfnamefont {J.~M.} \bibnamefont {Silver}}, \bibinfo {author} {\bibfnamefont {A.~{\O}.} \bibnamefont {Svela}}, \bibinfo {author} {\bibfnamefont {S.}~\bibnamefont {Zhang}},  and \bibinfo {author} {\bibfnamefont {P.}~\bibnamefont {Del'Haye}}, }\bibfield  {title} {\enquote {\bibinfo {title} {Logic gates based on interaction of counterpropagating light in microresonators},} }\href {http://opg.optica.org/jlt/abstract.cfm?URI=jlt-38-6-1414} {\bibfield  {journal} {\bibinfo  {journal} {J. Lightwave Technol.} }\textbf {\bibinfo {volume} {38}}, \bibinfo {pages} {1414--1419} (\bibinfo {year} {2020})}\BibitemShut {NoStop}%
\bibitem [{\citenamefont {White} \emph {et~al.}(2023)\citenamefont {White}, \citenamefont {Ahn}, \citenamefont {Gasse}, \citenamefont {Yang}, \citenamefont {Chang}, \citenamefont {Bowers}, and \citenamefont {Vu{\v{c}}kovi{\'c}}}]{white2023integrated}%
  \BibitemOpen
  \bibfield  {author} {\bibinfo {author} {\bibfnamefont {A.~D.} \bibnamefont {White}}, \bibinfo {author} {\bibfnamefont {G.~H.} \bibnamefont {Ahn}}, \bibinfo {author} {\bibfnamefont {K.~V.} \bibnamefont {Gasse}}, \bibinfo {author} {\bibfnamefont {K.~Y.} \bibnamefont {Yang}}, \bibinfo {author} {\bibfnamefont {L.}~\bibnamefont {Chang}}, \bibinfo {author} {\bibfnamefont {J.~E.} \bibnamefont {Bowers}},  and \bibinfo {author} {\bibfnamefont {J.}~\bibnamefont {Vu{\v{c}}kovi{\'c}}}, }\bibfield  {title} {\enquote {\bibinfo {title} {Integrated passive nonlinear optical isolators},} }\href@noop {} {\bibfield  {journal} {\bibinfo  {journal} {Nat. Photonics} }\textbf {\bibinfo {volume} {17}}, \bibinfo {pages} {143--149} (\bibinfo {year} {2023})}\BibitemShut {NoStop}%
\bibitem [{\citenamefont {Moroney} \emph {et~al.}(2022)\citenamefont {Moroney}, \citenamefont {Del~Bino}, \citenamefont {Zhang}, \citenamefont {Woodley}, \citenamefont {Hill}, \citenamefont {Wildi}, \citenamefont {Wittwer}, \citenamefont {Südmeyer}, \citenamefont {Oppo}, \citenamefont {Vanner}, \citenamefont {Brasch}, \citenamefont {Herr}, and \citenamefont {Del'Haye}}]{Moroney2022}%
  \BibitemOpen
  \bibfield  {author} {\bibinfo {author} {\bibfnamefont {N.}~\bibnamefont {Moroney}}, \bibinfo {author} {\bibfnamefont {L.}~\bibnamefont {Del~Bino}}, \bibinfo {author} {\bibfnamefont {S.}~\bibnamefont {Zhang}}, \bibinfo {author} {\bibfnamefont {M.~T.~M.} \bibnamefont {Woodley}}, \bibinfo {author} {\bibfnamefont {L.}~\bibnamefont {Hill}}, \bibinfo {author} {\bibfnamefont {T.}~\bibnamefont {Wildi}}, \bibinfo {author} {\bibfnamefont {V.~J.} \bibnamefont {Wittwer}}, \bibinfo {author} {\bibfnamefont {T.}~\bibnamefont {Südmeyer}}, \bibinfo {author} {\bibfnamefont {G.-L.} \bibnamefont {Oppo}}, \bibinfo {author} {\bibfnamefont {M.~R.} \bibnamefont {Vanner}}, \bibinfo {author} {\bibfnamefont {V.}~\bibnamefont {Brasch}}, \bibinfo {author} {\bibfnamefont {T.}~\bibnamefont {Herr}},  and \bibinfo {author} {\bibfnamefont {P.}~\bibnamefont {Del'Haye}}, }\bibfield  {title} {\enquote {\bibinfo {title} {A {K}err polarization controller},} }\href {\doibase 10.1038/s41467-021-27933-x} {\bibfield  {journal} {\bibinfo  {journal}
  {Nat. Commun.} }\textbf {\bibinfo {volume} {13}}, \bibinfo {pages} {398} (\bibinfo {year} {2022})}\BibitemShut {NoStop}%
\bibitem [{\citenamefont {Del~Bino} \emph {et~al.}(2021)\citenamefont {Del~Bino}, \citenamefont {Moroney}, and \citenamefont {Del’Haye}}]{del2021optical}%
  \BibitemOpen
  \bibfield  {author} {\bibinfo {author} {\bibfnamefont {L.}~\bibnamefont {Del~Bino}}, \bibinfo {author} {\bibfnamefont {N.}~\bibnamefont {Moroney}},  and \bibinfo {author} {\bibfnamefont {P.}~\bibnamefont {Del’Haye}}, }\bibfield  {title} {\enquote {\bibinfo {title} {Optical memories and switching dynamics of counterpropagating light states in microresonators},} }\href@noop {} {\bibfield  {journal} {\bibinfo  {journal} {Opt. Express} }\textbf {\bibinfo {volume} {29}}, \bibinfo {pages} {2193--2203} (\bibinfo {year} {2021})}\BibitemShut {NoStop}%
\bibitem [{\citenamefont {Yoshiki} and \citenamefont {Tanabe}(2014)}]{yoshiki2014all}%
  \BibitemOpen
  \bibfield  {author} {\bibinfo {author} {\bibfnamefont {W.}~\bibnamefont {Yoshiki}} and \bibinfo {author} {\bibfnamefont {T.}~\bibnamefont {Tanabe}}, }\bibfield  {title} {\enquote {\bibinfo {title} {All-optical switching using {K}err effect in a silica toroid microcavity},} }\href@noop {} {\bibfield  {journal} {\bibinfo  {journal} {Opt. Express} }\textbf {\bibinfo {volume} {22}}, \bibinfo {pages} {24332--24341} (\bibinfo {year} {2014})}\BibitemShut {NoStop}%
\bibitem [{\citenamefont {Hill} \emph {et~al.}(2023)\citenamefont {Hill}, \citenamefont {Oppo}, and \citenamefont {Del’Haye}}]{hill2023multi}%
  \BibitemOpen
  \bibfield  {author} {\bibinfo {author} {\bibfnamefont {L.}~\bibnamefont {Hill}}, \bibinfo {author} {\bibfnamefont {G.-L.} \bibnamefont {Oppo}},  and \bibinfo {author} {\bibfnamefont {P.}~\bibnamefont {Del’Haye}}, }\bibfield  {title} {\enquote {\bibinfo {title} {Multi-stage spontaneous symmetry breaking of light in {K}err ring resonators},} }\href@noop {} {\bibfield  {journal} {\bibinfo  {journal} {Commun. Phys.} }\textbf {\bibinfo {volume} {6}}, \bibinfo {pages} {208} (\bibinfo {year} {2023})}\BibitemShut {NoStop}%
\bibitem [{\citenamefont {Ghosh} \emph {et~al.}(2023)\citenamefont {Ghosh}, \citenamefont {Hill}, \citenamefont {Oppo}, and \citenamefont {Del'Haye}}]{ghosh2023four}%
  \BibitemOpen
  \bibfield  {author} {\bibinfo {author} {\bibfnamefont {A.}~\bibnamefont {Ghosh}}, \bibinfo {author} {\bibfnamefont {L.}~\bibnamefont {Hill}}, \bibinfo {author} {\bibfnamefont {G.-L.} \bibnamefont {Oppo}},  and \bibinfo {author} {\bibfnamefont {P.}~\bibnamefont {Del'Haye}}, }\bibfield  {title} {\enquote {\bibinfo {title} {Four-field symmetry breakings in twin-resonator photonic isomers},} }\href@noop {} {\bibfield  {journal} {\bibinfo  {journal} {Phys. Rev. Res.} }\textbf {\bibinfo {volume} {5}}, \bibinfo {pages} {L042012} (\bibinfo {year} {2023})}\BibitemShut {NoStop}%
\bibitem [{\citenamefont {Ghosh} \emph {et~al.}(2024)\citenamefont {Ghosh}, \citenamefont {Pal}, \citenamefont {Hill}, \citenamefont {Campbell}, \citenamefont {Bi}, \citenamefont {Zhang}, \citenamefont {Alabbadi}, \citenamefont {Zhang}, \citenamefont {Oppo}, and \citenamefont {Del'Haye}}]{ghosh2024controlled}%
  \BibitemOpen
  \bibfield  {author} {\bibinfo {author} {\bibfnamefont {A.}~\bibnamefont {Ghosh}}, \bibinfo {author} {\bibfnamefont {A.}~\bibnamefont {Pal}}, \bibinfo {author} {\bibfnamefont {L.}~\bibnamefont {Hill}}, \bibinfo {author} {\bibfnamefont {G.~N.} \bibnamefont {Campbell}}, \bibinfo {author} {\bibfnamefont {T.}~\bibnamefont {Bi}}, \bibinfo {author} {\bibfnamefont {Y.}~\bibnamefont {Zhang}}, \bibinfo {author} {\bibfnamefont {A.}~\bibnamefont {Alabbadi}}, \bibinfo {author} {\bibfnamefont {S.}~\bibnamefont {Zhang}}, \bibinfo {author} {\bibfnamefont {G.-L.} \bibnamefont {Oppo}},  and \bibinfo {author} {\bibfnamefont {P.}~\bibnamefont {Del'Haye}}, }\bibfield  {title} {\enquote {\bibinfo {title} {Controlled light distribution with coupled microresonator chains via {K}err symmetry breaking},} }\href@noop {} {\bibfield  {journal} {\bibinfo  {journal} {arXiv preprint arXiv:2402.10673} } (\bibinfo {year} {2024})}\BibitemShut {NoStop}%
\bibitem [{\citenamefont {Pal} \emph {et~al.}(2024)\citenamefont {Pal}, \citenamefont {Ghosh}, \citenamefont {Zhang}, \citenamefont {Hill}, \citenamefont {Yan}, \citenamefont {Zhang}, \citenamefont {Bi}, \citenamefont {Alabbadi}, and \citenamefont {Del'Haye}}]{pal2024linear}%
  \BibitemOpen
  \bibfield  {author} {\bibinfo {author} {\bibfnamefont {A.}~\bibnamefont {Pal}}, \bibinfo {author} {\bibfnamefont {A.}~\bibnamefont {Ghosh}}, \bibinfo {author} {\bibfnamefont {S.}~\bibnamefont {Zhang}}, \bibinfo {author} {\bibfnamefont {L.}~\bibnamefont {Hill}}, \bibinfo {author} {\bibfnamefont {H.}~\bibnamefont {Yan}}, \bibinfo {author} {\bibfnamefont {H.}~\bibnamefont {Zhang}}, \bibinfo {author} {\bibfnamefont {T.}~\bibnamefont {Bi}}, \bibinfo {author} {\bibfnamefont {A.}~\bibnamefont {Alabbadi}},  and \bibinfo {author} {\bibfnamefont {P.}~\bibnamefont {Del'Haye}}, }\bibfield  {title} {\enquote {\bibinfo {title} {Linear and nonlinear coupling of twin-resonators with {K}err nonlinearity},} }\href@noop {} {\bibfield  {journal} {\bibinfo  {journal} {arXiv preprint arXiv:2404.05646} } (\bibinfo {year} {2024})}\BibitemShut {NoStop}%
\bibitem [{\citenamefont {Hill} \emph {et~al.}(2020)\citenamefont {Hill}, \citenamefont {Oppo}, \citenamefont {Woodley}, and \citenamefont {Del'Haye}}]{hill2020effects}%
  \BibitemOpen
  \bibfield  {author} {\bibinfo {author} {\bibfnamefont {L.}~\bibnamefont {Hill}}, \bibinfo {author} {\bibfnamefont {G.-L.} \bibnamefont {Oppo}}, \bibinfo {author} {\bibfnamefont {M.~T.~M.} \bibnamefont {Woodley}},  and \bibinfo {author} {\bibfnamefont {P.}~\bibnamefont {Del'Haye}}, }\bibfield  {title} {\enquote {\bibinfo {title} {Effects of self- and cross-phase modulation on the spontaneous symmetry breaking of light in ring resonators},} }\href {\doibase 10.1103/PhysRevA.101.013823} {\bibfield  {journal} {\bibinfo  {journal} {Phys. Rev. A} }\textbf {\bibinfo {volume} {101}}, \bibinfo {pages} {013823} (\bibinfo {year} {2020})}\BibitemShut {NoStop}%
\bibitem [{\citenamefont {Del'Haye} \emph {et~al.}(2013)\citenamefont {Del'Haye}, \citenamefont {Diddams}, and \citenamefont {Papp}}]{del2013laser}%
  \BibitemOpen
  \bibfield  {author} {\bibinfo {author} {\bibfnamefont {P.}~\bibnamefont {Del'Haye}}, \bibinfo {author} {\bibfnamefont {S.~A.} \bibnamefont {Diddams}},  and \bibinfo {author} {\bibfnamefont {S.~B.} \bibnamefont {Papp}}, }\bibfield  {title} {\enquote {\bibinfo {title} {Laser-machined ultra-high-{Q} microrod resonators for nonlinear optics},} }\href@noop {} {\bibfield  {journal} {\bibinfo  {journal} {Appl. Phys. Lett.} }\textbf {\bibinfo {volume} {102}} (\bibinfo {year} {2013})}\BibitemShut {NoStop}%
\bibitem [{\citenamefont {Lugiato} and \citenamefont {Lefever}(1987)}]{lugiato1987spatial}%
  \BibitemOpen
  \bibfield  {author} {\bibinfo {author} {\bibfnamefont {L.~A.} \bibnamefont {Lugiato}} and \bibinfo {author} {\bibfnamefont {R.}~\bibnamefont {Lefever}}, }\bibfield  {title} {\enquote {\bibinfo {title} {Spatial dissipative structures in passive optical systems},} }\href@noop {} {\bibfield  {journal} {\bibinfo  {journal} {Phys. Rev. Lett.} }\textbf {\bibinfo {volume} {58}}, \bibinfo {pages} {2209} (\bibinfo {year} {1987})}\BibitemShut {NoStop}%
\bibitem [{\citenamefont {Riemensberger} \emph {et~al.}(2022)\citenamefont {Riemensberger}, \citenamefont {Kuznetsov}, \citenamefont {Liu}, \citenamefont {He}, \citenamefont {Wang}, and \citenamefont {Kippenberg}}]{riemensberger2022photonic}%
  \BibitemOpen
  \bibfield  {author} {\bibinfo {author} {\bibfnamefont {J.}~\bibnamefont {Riemensberger}}, \bibinfo {author} {\bibfnamefont {N.}~\bibnamefont {Kuznetsov}}, \bibinfo {author} {\bibfnamefont {J.}~\bibnamefont {Liu}}, \bibinfo {author} {\bibfnamefont {J.}~\bibnamefont {He}}, \bibinfo {author} {\bibfnamefont {R.~N.} \bibnamefont {Wang}},  and \bibinfo {author} {\bibfnamefont {T.~J.} \bibnamefont {Kippenberg}}, }\bibfield  {title} {\enquote {\bibinfo {title} {A photonic integrated continuous-travelling-wave parametric amplifier},} }\href@noop {} {\bibfield  {journal} {\bibinfo  {journal} {Nature} }\textbf {\bibinfo {volume} {612}}, \bibinfo {pages} {56--61} (\bibinfo {year} {2022})}\BibitemShut {NoStop}%
\bibitem [{\citenamefont {Zhang} \emph {et~al.}(2024)\citenamefont {Zhang}, \citenamefont {Bi}, \citenamefont {Harder}, \citenamefont {Ohletz}, \citenamefont {Gannott}, \citenamefont {Gumann}, \citenamefont {Butzen}, \citenamefont {Zhang}, and \citenamefont {Del'Haye}}]{zhang2024low}%
  \BibitemOpen
  \bibfield  {author} {\bibinfo {author} {\bibfnamefont {S.}~\bibnamefont {Zhang}}, \bibinfo {author} {\bibfnamefont {T.}~\bibnamefont {Bi}}, \bibinfo {author} {\bibfnamefont {I.}~\bibnamefont {Harder}}, \bibinfo {author} {\bibfnamefont {O.}~\bibnamefont {Ohletz}}, \bibinfo {author} {\bibfnamefont {F.}~\bibnamefont {Gannott}}, \bibinfo {author} {\bibfnamefont {A.}~\bibnamefont {Gumann}}, \bibinfo {author} {\bibfnamefont {E.}~\bibnamefont {Butzen}}, \bibinfo {author} {\bibfnamefont {Y.}~\bibnamefont {Zhang}},  and \bibinfo {author} {\bibfnamefont {P.}~\bibnamefont {Del'Haye}}, }\bibfield  {title} {\enquote {\bibinfo {title} {Low-temperature sputtered ultralow-loss silicon nitride for hybrid photonic integration},} }\href@noop {} {\bibfield  {journal} {\bibinfo  {journal} {Laser \& Photonics Reviews} }\textbf {\bibinfo {volume} {18}}, \bibinfo {pages} {2300642} (\bibinfo {year} {2024})}\BibitemShut {NoStop}%
\bibitem [{\citenamefont {Chiles} \emph {et~al.}(2018)\citenamefont {Chiles}, \citenamefont {Nader}, \citenamefont {Hickstein}, \citenamefont {Yu}, \citenamefont {Briles}, \citenamefont {Carlson}, \citenamefont {Jung}, \citenamefont {Shainline}, \citenamefont {Diddams}, \citenamefont {Papp} \emph {et~al.}}]{chiles2018deuterated}%
  \BibitemOpen
  \bibfield  {author} {\bibinfo {author} {\bibfnamefont {J.}~\bibnamefont {Chiles}}, \bibinfo {author} {\bibfnamefont {N.}~\bibnamefont {Nader}}, \bibinfo {author} {\bibfnamefont {D.~D.} \bibnamefont {Hickstein}}, \bibinfo {author} {\bibfnamefont {S.~P.} \bibnamefont {Yu}}, \bibinfo {author} {\bibfnamefont {T.~C.} \bibnamefont {Briles}}, \bibinfo {author} {\bibfnamefont {D.}~\bibnamefont {Carlson}}, \bibinfo {author} {\bibfnamefont {H.}~\bibnamefont {Jung}}, \bibinfo {author} {\bibfnamefont {J.~M.} \bibnamefont {Shainline}}, \bibinfo {author} {\bibfnamefont {S.}~\bibnamefont {Diddams}}, \bibinfo {author} {\bibfnamefont {S.~B.} \bibnamefont {Papp}},  \emph {et~al.}, }\bibfield  {title} {\enquote {\bibinfo {title} {Deuterated silicon nitride photonic devices for broadband optical frequency comb generation},} }\href@noop {} {\bibfield  {journal} {\bibinfo  {journal} {Opt. Lett.} }\textbf {\bibinfo {volume} {43}}, \bibinfo {pages} {1527--1530} (\bibinfo {year} {2018})}\BibitemShut {NoStop}%
\bibitem [{\citenamefont {Pal} \emph {et~al.}(2023)\citenamefont {Pal}, \citenamefont {Ghosh}, \citenamefont {Zhang}, \citenamefont {Bi}, and \citenamefont {Del’Haye}}]{pal2023machine}%
  \BibitemOpen
  \bibfield  {author} {\bibinfo {author} {\bibfnamefont {A.}~\bibnamefont {Pal}}, \bibinfo {author} {\bibfnamefont {A.}~\bibnamefont {Ghosh}}, \bibinfo {author} {\bibfnamefont {S.}~\bibnamefont {Zhang}}, \bibinfo {author} {\bibfnamefont {T.}~\bibnamefont {Bi}},  and \bibinfo {author} {\bibfnamefont {P.}~\bibnamefont {Del’Haye}}, }\bibfield  {title} {\enquote {\bibinfo {title} {Machine learning assisted inverse design of microresonators},} }\href@noop {} {\bibfield  {journal} {\bibinfo  {journal} {Opt. Express} }\textbf {\bibinfo {volume} {31}}, \bibinfo {pages} {8020--8028} (\bibinfo {year} {2023})}\BibitemShut {NoStop}%
\bibitem [{\citenamefont {Li} \emph {et~al.}(2020)\citenamefont {Li}, \citenamefont {Huang}, \citenamefont {Li}, \citenamefont {Liu}, \citenamefont {Yang}, \citenamefont {Vinod}, \citenamefont {Wang}, \citenamefont {Yu}, \citenamefont {Kwong}, \citenamefont {Wang} \emph {et~al.}}]{li2020real}%
  \BibitemOpen
  \bibfield  {author} {\bibinfo {author} {\bibfnamefont {Y.}~\bibnamefont {Li}}, \bibinfo {author} {\bibfnamefont {S.-W.} \bibnamefont {Huang}}, \bibinfo {author} {\bibfnamefont {B.}~\bibnamefont {Li}}, \bibinfo {author} {\bibfnamefont {H.}~\bibnamefont {Liu}}, \bibinfo {author} {\bibfnamefont {J.}~\bibnamefont {Yang}}, \bibinfo {author} {\bibfnamefont {A.~K.} \bibnamefont {Vinod}}, \bibinfo {author} {\bibfnamefont {K.}~\bibnamefont {Wang}}, \bibinfo {author} {\bibfnamefont {M.}~\bibnamefont {Yu}}, \bibinfo {author} {\bibfnamefont {D.-L.} \bibnamefont {Kwong}}, \bibinfo {author} {\bibfnamefont {H.-T.} \bibnamefont {Wang}},  \emph {et~al.}, }\bibfield  {title} {\enquote {\bibinfo {title} {Real-time transition dynamics and stability of chip-scale dispersion-managed frequency microcombs},} }\href@noop {} {\bibfield  {journal} {\bibinfo  {journal} {Light Sci. Appl.} }\textbf {\bibinfo {volume} {9}}, \bibinfo {pages} {52} (\bibinfo {year} {2020})}\BibitemShut {NoStop}%
\bibitem [{\citenamefont {Fujii} and \citenamefont {Tanabe}(2020)}]{fujii2020dispersion}%
  \BibitemOpen
  \bibfield  {author} {\bibinfo {author} {\bibfnamefont {S.}~\bibnamefont {Fujii}} and \bibinfo {author} {\bibfnamefont {T.}~\bibnamefont {Tanabe}}, }\bibfield  {title} {\enquote {\bibinfo {title} {Dispersion engineering and measurement of whispering gallery mode microresonator for {K}err frequency comb generation},} }\href@noop {} {\bibfield  {journal} {\bibinfo  {journal} {Nanophotonics} }\textbf {\bibinfo {volume} {9}}, \bibinfo {pages} {1087--1104} (\bibinfo {year} {2020})}\BibitemShut {NoStop}%
\bibitem [{\citenamefont {Bi} \emph {et~al.}(2023)\citenamefont {Bi}, \citenamefont {Zhang}, \citenamefont {Ghosh}, \citenamefont {Lohse}, \citenamefont {Harder}, \citenamefont {Yang}, and \citenamefont {Del'Haye}}]{bi2023chip}%
  \BibitemOpen
  \bibfield  {author} {\bibinfo {author} {\bibfnamefont {T.}~\bibnamefont {Bi}}, \bibinfo {author} {\bibfnamefont {S.}~\bibnamefont {Zhang}}, \bibinfo {author} {\bibfnamefont {A.}~\bibnamefont {Ghosh}}, \bibinfo {author} {\bibfnamefont {O.}~\bibnamefont {Lohse}}, \bibinfo {author} {\bibfnamefont {I.}~\bibnamefont {Harder}}, \bibinfo {author} {\bibfnamefont {K.~Y.} \bibnamefont {Yang}},  and \bibinfo {author} {\bibfnamefont {P.}~\bibnamefont {Del'Haye}}, }\bibfield  {title} {\enquote {\bibinfo {title} {On-chip inverse designed {F}abry-{P}{\'e}rot resonators},} }in \href@noop {} {\emph {\bibinfo {booktitle} {2023 Conference on Lasers and Electro-Optics Europe \& European Quantum Electronics Conference (CLEO/Europe-EQEC)}}} (\bibinfo {organization} {IEEE}, \bibinfo {year} {2023}) pp. \bibinfo {pages} {1--1}\BibitemShut {NoStop}%
\bibitem [{\citenamefont {Xu} \emph {et~al.}(2021)\citenamefont {Xu}, \citenamefont {Nielsen}, \citenamefont {Garbin}, \citenamefont {Hill}, \citenamefont {Oppo}, \citenamefont {Fatome}, \citenamefont {Murdoch}, \citenamefont {Coen}, and \citenamefont {Erkintalo}}]{Xu2021}%
  \BibitemOpen
  \bibfield  {author} {\bibinfo {author} {\bibfnamefont {G.}~\bibnamefont {Xu}}, \bibinfo {author} {\bibfnamefont {A.~U.} \bibnamefont {Nielsen}}, \bibinfo {author} {\bibfnamefont {B.}~\bibnamefont {Garbin}}, \bibinfo {author} {\bibfnamefont {L.}~\bibnamefont {Hill}}, \bibinfo {author} {\bibfnamefont {G.-L.} \bibnamefont {Oppo}}, \bibinfo {author} {\bibfnamefont {J.}~\bibnamefont {Fatome}}, \bibinfo {author} {\bibfnamefont {S.~G.} \bibnamefont {Murdoch}}, \bibinfo {author} {\bibfnamefont {S.}~\bibnamefont {Coen}},  and \bibinfo {author} {\bibfnamefont {M.}~\bibnamefont {Erkintalo}}, }\bibfield  {title} {\enquote {\bibinfo {title} {Spontaneous symmetry breaking of dissipative optical solitons in a two-component {K}err resonator},} }\href {\doibase 10.1038/s41467-021-24251-0} {\bibfield  {journal} {\bibinfo  {journal} {Nat. Commun.} }\textbf {\bibinfo {volume} {12}}, \bibinfo {pages} {4023} (\bibinfo {year} {2021})}\BibitemShut {NoStop}%
\bibitem [{\citenamefont {Xu} \emph {et~al.}(2022)\citenamefont {Xu}, \citenamefont {Hill}, \citenamefont {Fatome}, \citenamefont {Oppo}, \citenamefont {Erkintalo}, \citenamefont {Murdoch}, and \citenamefont {Coen}}]{Xu:22}%
  \BibitemOpen
  \bibfield  {author} {\bibinfo {author} {\bibfnamefont {G.}~\bibnamefont {Xu}}, \bibinfo {author} {\bibfnamefont {L.}~\bibnamefont {Hill}}, \bibinfo {author} {\bibfnamefont {J.}~\bibnamefont {Fatome}}, \bibinfo {author} {\bibfnamefont {G.-L.} \bibnamefont {Oppo}}, \bibinfo {author} {\bibfnamefont {M.}~\bibnamefont {Erkintalo}}, \bibinfo {author} {\bibfnamefont {S.~G.} \bibnamefont {Murdoch}},  and \bibinfo {author} {\bibfnamefont {S.}~\bibnamefont {Coen}}, }\bibfield  {title} {\enquote {\bibinfo {title} {Breathing dynamics of symmetry-broken temporal cavity solitons in {K}err ring resonators},} }\href {\doibase 10.1364/OL.449679} {\bibfield  {journal} {\bibinfo  {journal} {Opt. Lett.} }\textbf {\bibinfo {volume} {47}}, \bibinfo {pages} {1486--1489} (\bibinfo {year} {2022})}\BibitemShut {NoStop}%
\bibitem [{\citenamefont {Yan} \emph {et~al.}(2024)\citenamefont {Yan}, \citenamefont {Ghosh}, \citenamefont {Pal}, \citenamefont {Zhang}, \citenamefont {Bi}, \citenamefont {Ghalanos}, \citenamefont {Zhang}, \citenamefont {Hill}, \citenamefont {Zhang}, \citenamefont {Zhuang} \emph {et~al.}}]{yan2024real}%
  \BibitemOpen
  \bibfield  {author} {\bibinfo {author} {\bibfnamefont {H.}~\bibnamefont {Yan}}, \bibinfo {author} {\bibfnamefont {A.}~\bibnamefont {Ghosh}}, \bibinfo {author} {\bibfnamefont {A.}~\bibnamefont {Pal}}, \bibinfo {author} {\bibfnamefont {H.}~\bibnamefont {Zhang}}, \bibinfo {author} {\bibfnamefont {T.}~\bibnamefont {Bi}}, \bibinfo {author} {\bibfnamefont {G.}~\bibnamefont {Ghalanos}}, \bibinfo {author} {\bibfnamefont {S.}~\bibnamefont {Zhang}}, \bibinfo {author} {\bibfnamefont {L.}~\bibnamefont {Hill}}, \bibinfo {author} {\bibfnamefont {Y.}~\bibnamefont {Zhang}}, \bibinfo {author} {\bibfnamefont {Y.}~\bibnamefont {Zhuang}},  \emph {et~al.}, }\bibfield  {title} {\enquote {\bibinfo {title} {Real-time imaging of standing-wave patterns in microresonators},} }\href@noop {} {\bibfield  {journal} {\bibinfo  {journal} {Proc. Natl. Acad. Sci. U.S.A.} }\textbf {\bibinfo {volume} {121}}, \bibinfo {pages} {e2313981121} (\bibinfo {year} {2024})}\BibitemShut {NoStop}%
\bibitem [{\citenamefont {Anashkina} and \citenamefont {Andrianov}(2024)}]{anashkina2024phase}%
  \BibitemOpen
  \bibfield  {author} {\bibinfo {author} {\bibfnamefont {E.~A.} \bibnamefont {Anashkina}} and \bibinfo {author} {\bibfnamefont {A.~V.} \bibnamefont {Andrianov}}, }\bibfield  {title} {\enquote {\bibinfo {title} {Phase-sensitive symmetry breaking in bidirectionally pumped {K}err microresonators},} }\href@noop {} {\bibfield  {journal} {\bibinfo  {journal} {arXiv preprint arXiv:2407.07594} } (\bibinfo {year} {2024})}\BibitemShut {NoStop}%
\bibitem [{\citenamefont {Woodley} \emph {et~al.}(2021)\citenamefont {Woodley}, \citenamefont {Hill}, \citenamefont {Del~Bino}, \citenamefont {Oppo}, and \citenamefont {Del’Haye}}]{woodley2021self}%
  \BibitemOpen
  \bibfield  {author} {\bibinfo {author} {\bibfnamefont {M.~T.} \bibnamefont {Woodley}}, \bibinfo {author} {\bibfnamefont {L.}~\bibnamefont {Hill}}, \bibinfo {author} {\bibfnamefont {L.}~\bibnamefont {Del~Bino}}, \bibinfo {author} {\bibfnamefont {G.-L.} \bibnamefont {Oppo}},  and \bibinfo {author} {\bibfnamefont {P.}~\bibnamefont {Del’Haye}}, }\bibfield  {title} {\enquote {\bibinfo {title} {Self-switching {K}err oscillations of counterpropagating light in microresonators},} }\href@noop {} {\bibfield  {journal} {\bibinfo  {journal} {Physical Review Letters} }\textbf {\bibinfo {volume} {126}}, \bibinfo {pages} {043901} (\bibinfo {year} {2021})}\BibitemShut {NoStop}%
\bibitem [{\citenamefont {Tusnin} \emph {et~al.}(2023)\citenamefont {Tusnin}, \citenamefont {Tikan}, \citenamefont {Komagata}, and \citenamefont {Kippenberg}}]{tusnin2023nonlinear}%
  \BibitemOpen
  \bibfield  {author} {\bibinfo {author} {\bibfnamefont {A.}~\bibnamefont {Tusnin}}, \bibinfo {author} {\bibfnamefont {A.}~\bibnamefont {Tikan}}, \bibinfo {author} {\bibfnamefont {K.}~\bibnamefont {Komagata}},  and \bibinfo {author} {\bibfnamefont {T.~J.} \bibnamefont {Kippenberg}}, }\bibfield  {title} {\enquote {\bibinfo {title} {Nonlinear dynamics and {K}err frequency comb formation in lattices of coupled microresonators},} }\href@noop {} {\bibfield  {journal} {\bibinfo  {journal} {Commun. Phys.} }\textbf {\bibinfo {volume} {6}}, \bibinfo {pages} {317} (\bibinfo {year} {2023})}\BibitemShut {NoStop}%
\end{thebibliography}%

\onecolumngrid

\appendix

\clearpage
\section{\large{Appendix A: SSB of phases - Analytical Solutions}}
\label{A}
\begin{figure*}[b]
\includegraphics[width=0.6\columnwidth]{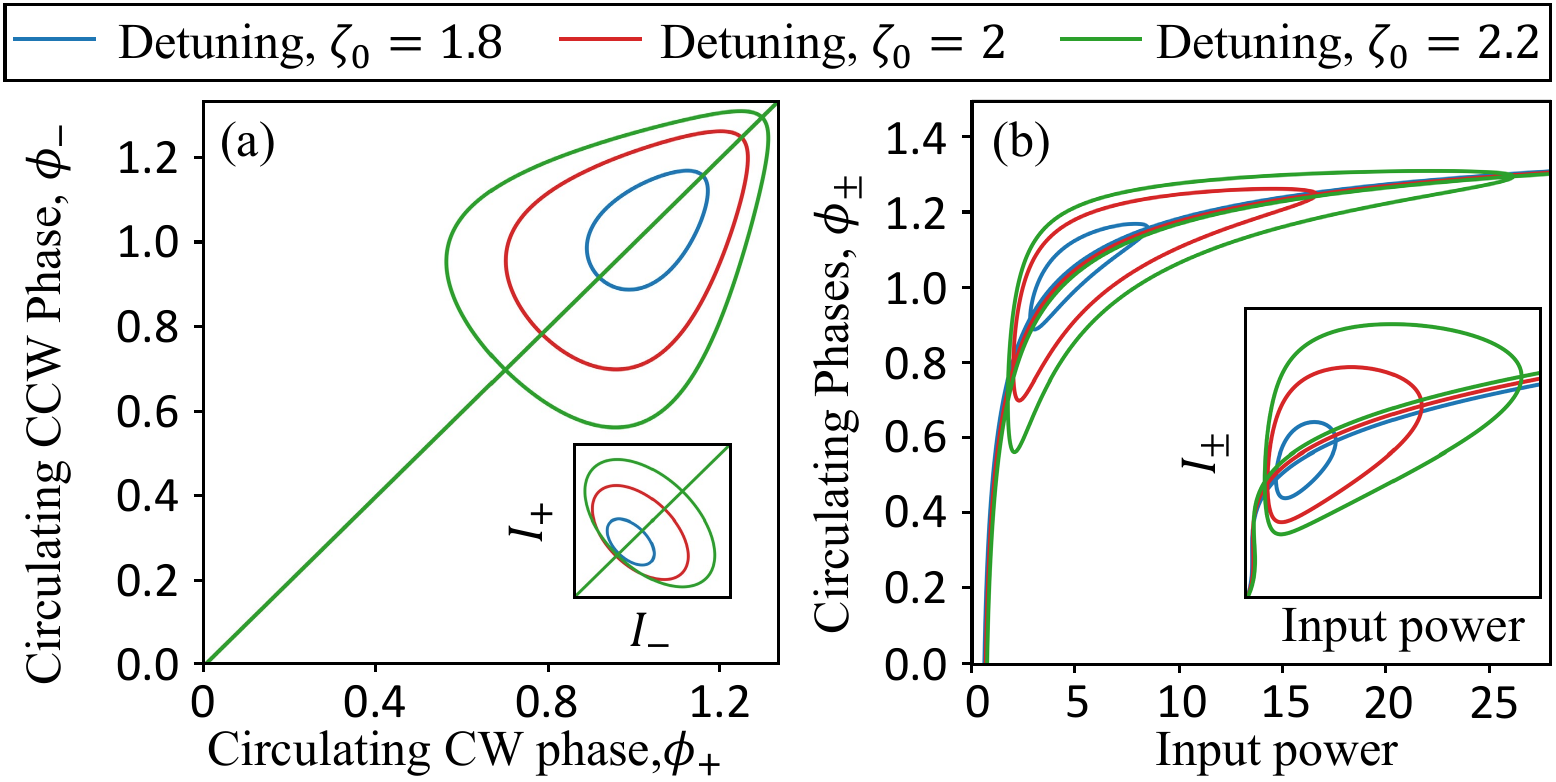}
\caption {\textit{Analytical solutions for symmetry breaking in phase.} Panel (a) shows the relationships between the phases of the two counterpropagating fields. The straight line represents the symmetric solutions, whereas the ellipses correspond to the asymmetric solutions for different detunings. The inset shows the corresponding relationships between the intensities of the two counterpropagating fields ($I_+$ and $I_-$). Panel (b) shows the phases as functions of input power with an inset illustrating the variation in the corresponding field intensities as a function of input power for different detuning values.
}
\label{PhaseSSBTheory}
\end{figure*}

The steady-state solutions of the system can be obtained by setting Eq.~(1) to $0$. Detailed calculations for obtaining the steady-state intensities have already been performed in Ref.~\cite{woodley2018universal}. The circulating intensities in the two directions follow the relation:
\begin{equation}
\left(I_+ - I_-\right)\left(I_{+}^2 + I_{-}^2 - 2\zeta_0 I_+ - 2\zeta_0 I_{-} + I_{+}I_{-} + 1+\zeta_0^2\right)=0,
\label{IntsSS}
\end{equation}
where, $I_{\pm} = |E_{\pm}|^2$. If we consider the circulating fields $E_{\pm}$ with corresponding phases $\phi_{\pm}$, i.e., $E_{\pm} = \sqrt{I_{\pm}}e^{i\phi_{\pm}}$, the steady-state response for the phases becomes,
\begin{equation}
\left(x-y\right)\left(x^2 - xy + y^2 - \zeta_0 x - \zeta_0 y +3\right)=0,
\label{PhasesSS}
\end{equation}
where $x= \tan(\phi_+)$ and $y= \tan(\phi_-)$. The expressions within the first set of round brackets on the LHS of both Eqs.~\eqref{IntsSS} and~\eqref{PhasesSS} represent the symmetric lines, and the second set of round brackets represent the Kerr-effect induced asymmetric ellipses, as seen in Fig.~\ref{PhaseSSBTheory}(a) where the two axes are the phases of the fields in each direction. The inset in Fig.~\ref{PhaseSSBTheory}(a) shows the intensity evolution of one circulating field with respect to its counter-propagating field. The points where the symmetric lines meet the respective asymmetric ellipses denote the endpoints of the SSB region and are given as:
\begin{eqnarray}
I_{\pm,0}=\frac{2\zeta_0 \pm \sqrt{\zeta_0^2 - 3}}{3},\\
\phi_{\pm,0} = \tan^{-1} \left(\zeta_0 \pm \sqrt{\zeta_0^2 - 3}\right).
\label{PhasesSSSol}
\end{eqnarray}

Figure~\ref{PhaseSSBTheory}(b) depicts the analytical solutions for the phases of the fields as a function of input power. For low input power, the phases of the counter-propagating fields remain symmetric. After a certain threshold of input power, the symmetric line becomes unstable (similar to the case of SSB in intensities~\cite{woodley2018universal}), and following a bifurcation point, the phases continue increasing but along a symmetry broken solution set. At the beginning of the SSB region, the phase of one field increases and the phase in the other direction decreases. This dominant direction is randomly chosen. Phases of both fields reach extrema, after which they approach each other again and finally converge in an inverse bifurcation point. The following symmetric solution line regains its stability after this inverse bifurcation point~\cite{woodley2018universal}. Similarly, intensities also become asymmetric following the bifurcation, as shown in the inset of Fig.~\ref{PhaseSSBTheory}(b). It should be noted that the SSB bifurcations occur at the same input power for both the phases and intensities of the circulating fields. With increasing detuning, the size of the bubble increases in Fig.~\ref{PhaseSSBTheory}(a,b).


\section{\large{Appendix B: Simulations - phase oscillations and bistability jumps}}
\label{B}
\begin{figure*}[h]
\includegraphics[width=0.6\textwidth]{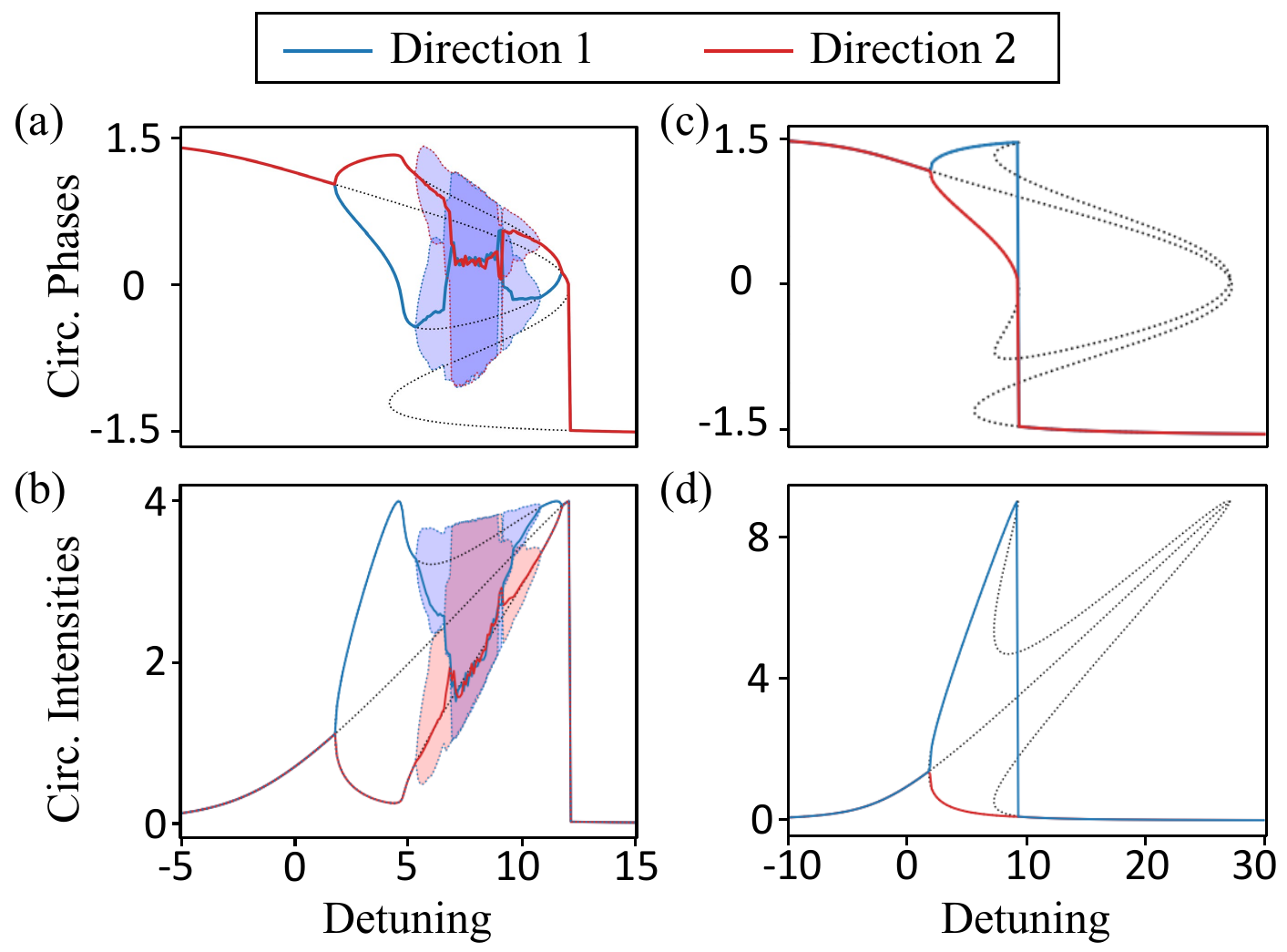}
\caption {\textit{Simulated solutions for symmetry breaking in phases for various input powers.} (a,b) With increasing input power from the value used in Fig.~4, both phases and intensities of the circulating fields show oscillations. The colored dots represent the maxima and minima of the respective fields. (c,d) For even higher input power, bistabilities appear within spontaneous symmetry breaking bubbles. The black dotted lines show analytical solutions, the red and blue solid lines represent the circulating fields. Used parameters: $|f|^2 = 4$~\cite{woodley2021self}. More details on the oscillations in the intensities can be found in Ref.~\cite{woodley2021self}.}
\label{PhaseSSBSimsSM}
\end{figure*}

The upper (lower) panel of Fig.~\ref{PhaseSSBSimsSM} shows detuning scans of phases (intensities) of the circulating fields in the two directions of a microresonator. 

For higher input power compared to the case of Fig.~\ref{PhaseSSBSimsSM}, we observe a horn-like shape in the SSB bubble (i.e., the SSB bubble starts to fold on itself). This happens both in phase (Fig.~\ref{PhaseSSBSimsSM}(a)) and intensity (Fig.~\ref{PhaseSSBSimsSM}(b)). Following the horn-like region in the bubble, the asymmetric branches enter an oscillating region. The oscillations in phases can are shown in Fig.~\ref{PhaseSSBSimsSM}(a) and happen simultaneously with the oscillations in intensities shown in Fig.~\ref{PhaseSSBSimsSM}(b). We report that oscillations in circulating field intensities in the SSB region are accompanied by oscillations in the phases of the fields. The colored dotted lines represent the maxima and minima of the oscillations for each detuning. It can be observed that with increasing detuning, the amplitudes of the phase oscillation increase, and after a certain detuning, the oscillations of the phases of the two fields overlap. The average phases, depicted in solid lines, regain symmetry in a small region within the oscillation region. The amplitudes of the phase oscillations reach maxima and then reduce with further increasing the detuning, causing another region of non-overapping oscillations before they cease to exist. These different types of oscillations in phases and the corresponding oscillations in intensities are depicted in the next section. After the oscillatory region, the phases of the fields follow the asymmetric branches and merge into a symmetric branch at the end of the SSB bubble. 

With further increasing the input power, the folds on the bubble turn sharper causing bistability within the bubble, as shown in Fig.~\ref{PhaseSSBSimsSM}(c) and (d). The bistability in intensity leads to a bistability jump inside the bubble, which is large enough to cause the intermediate oscillatory stages and lead to the fields jumping down to the stable symmetric lower branch. This bistability jump can be observed in the phases of the circulating fields as well, as shown in Fig.~\ref{PhaseSSBSimsSM}(c).

\clearpage
\section{\large{Appendix C: Examples of oscillations in phases}}
\label{C}
\begin{figure*}[h!]
\includegraphics[width=0.65\columnwidth]{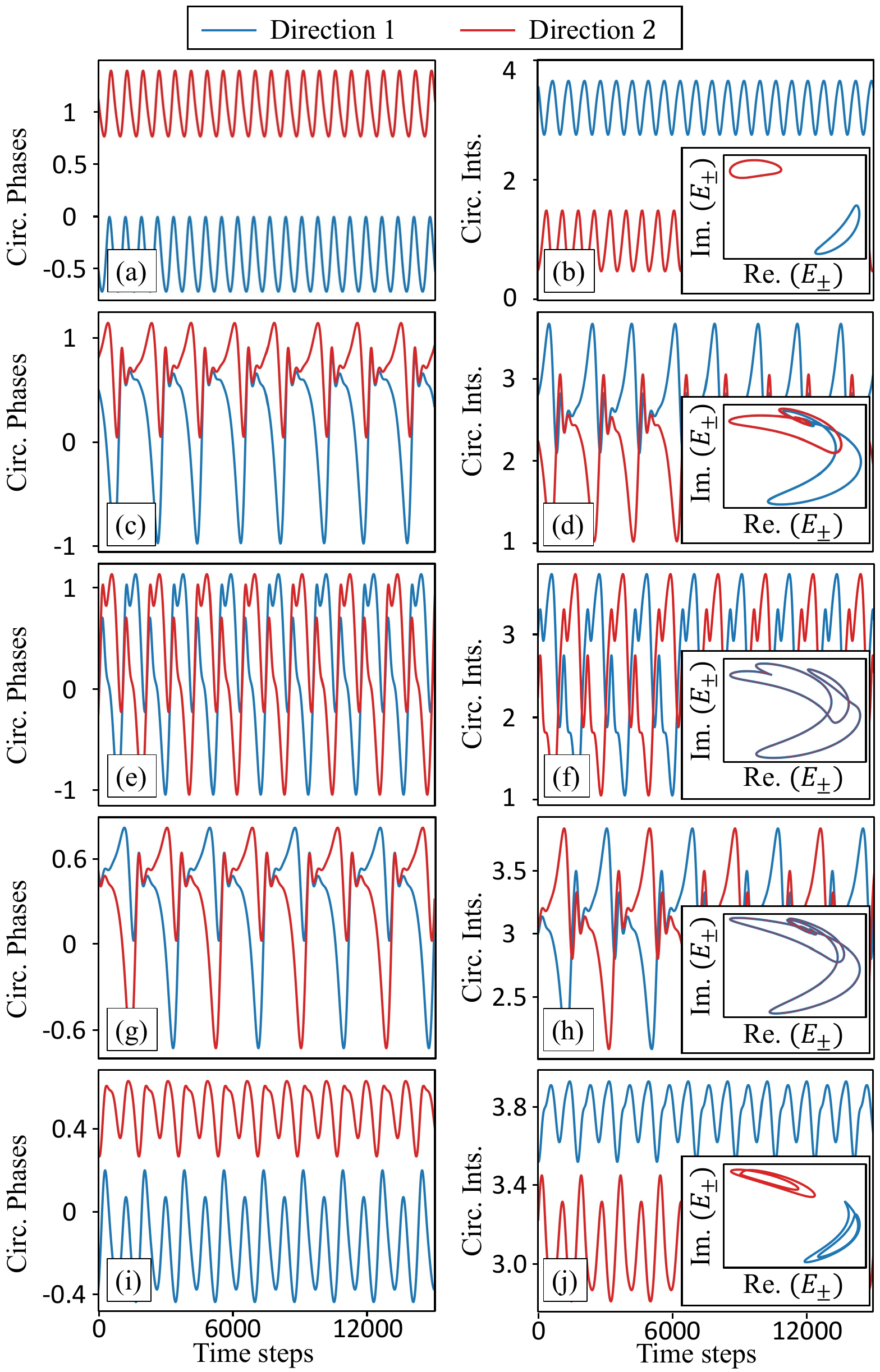}
\caption {\textit{Oscillations in phases and intensities under symmetry broken conditions observed in simulations.} The left (right) panel shows the temporal oscillations in the phases (intensities) of the fields for input power, $|f|^2 = 4$ and detuning values, $\zeta_0 =~5.5652~(\text{panels (a,b)}),~6.7826~(\text{panels (c,d)}),~7.0435~(\text{panels (e,f)}),~8.8696~(\text{panels (g,h)}),~\text{and }~10.087~(\text{panels (i,j)})$. The insets of the right panels show the real-vs-imaginary phase space diagrams of the circulating fields. With increasing detuning, the oscillations (in both phases and intensities) grow initially, leading to overlapping oscillations (panels (c,d)) and perfect periodic switchings (panels (e-h)). However, after a certain detuning, the oscillations become non-overlapping (panels (i,j)).}
\label{Oscillations_SM}
\end{figure*}

Figure~\ref{Oscillations_SM} shows examples of oscillations in phases and intensities of circulating counterpropagating fields in a microresonator that is pumped symmetrically and bidirectionally.

\clearpage
\section{\large{Appendix D: SSB of phases in coupled resonator systems}}
\label{D}
\begin{figure*}[h!]
\includegraphics[width=0.5\columnwidth]{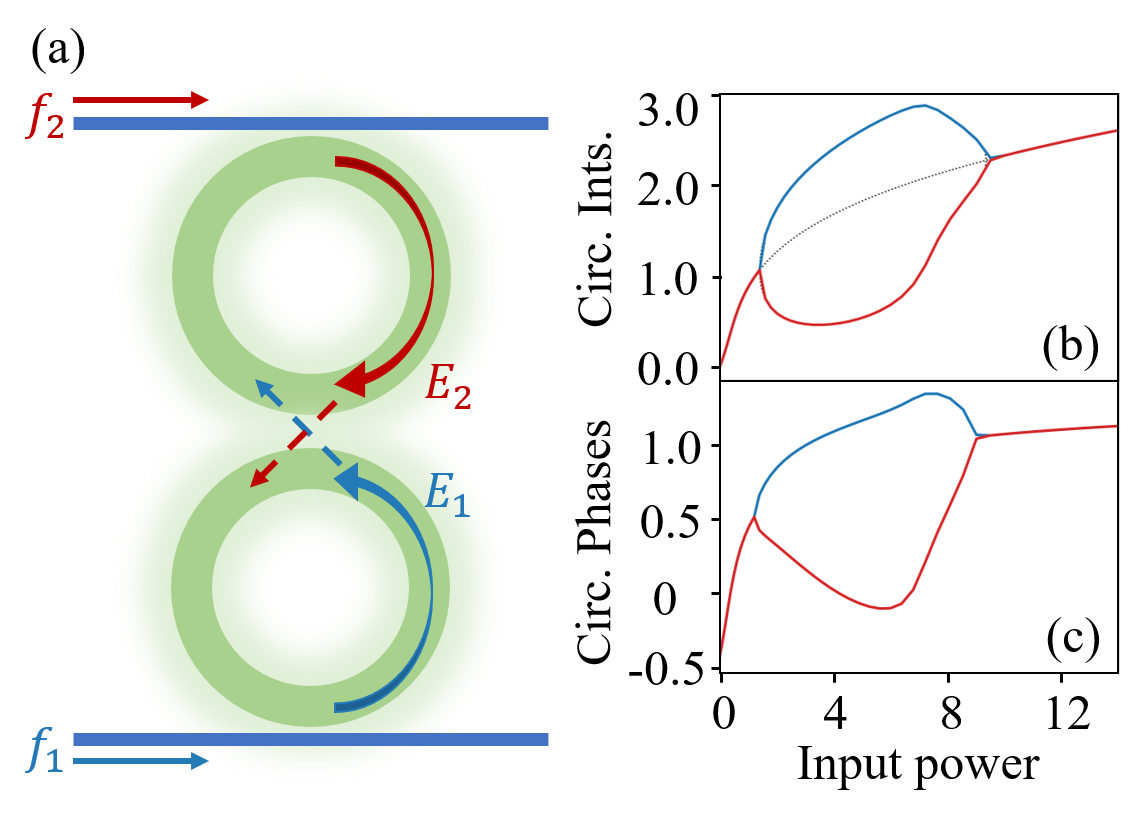}
\caption {\textit{Simulated solutions for symmetry breaking in intensities and phases in coupled resonator systems.} Panel (a) shows the schematics of a coupled resonator system, where both resonators are pumped in such a way that light fields are propagating in one certain direction in both resonators (clockwise in resonator 1 and counter-clockwise in resonator 2). Panels (b) and (c) depict the spontaneous symmetry breakings in intensities and phases of the circulating fields in the two resonators respectively. The dashed black line presents the unstable analytical solutions for the field intensities within the resonators.}
\label{CoupledResonators_SM}
\end{figure*}

In this section, we demonstrate that SSB between the circulating field intensities in the two coupled resonators are also accompanied by SSB in the phases of the fields. We model the evaluations of the fields in the coupled system depicted in Fig.~\ref{CoupledResonators_SM}(a) by a coupled LLE~\cite{lugiato1987spatial,tusnin2023nonlinear}
\begin{equation}
\frac{\partial E_{n}}{\partial t} = - E_{n} - i\zeta_0 E_{n} + ijE_{m} + i|E_{n}|^2E_{n} + f_n,
\label{CoupledLLEquations}
\end{equation}
where $E_n$ is the normalized optical field envelope for the $n^\text{th}$ resonator, $\zeta_0$ is the normalized detuning, $j$ is the normalized inter-resonator coupling rate and $f_n$ is the input to the $n^\text{th}$ resonator. Here, all the terms have been normalized with respect to half of the total loss of each resonator. We consider here that the system is perfectly symmetrical, therefore, $f_n = f$ is the same for both the resonators.

From Fig.~\ref{CoupledResonators_SM} (b) and (c), it can be stated that the phases of the circulating fields in the two resonators undergo SSB as well when there is an SSB in their corresponding intensities. Therefore, one can build all-optical switches or optical digital logic gates using these resonators as well.

\end{document}